\documentclass[aps,onecolumn,12pt,showpacs,preprintnumbers]{revtex4}

\usepackage{amsmath}
\usepackage{bm}
\usepackage{graphicx}
\usepackage{epstopdf}

\begin{document}

\title{Axion cold dark matter in non-standard cosmologies}

\author{Luca Visinelli}
\email{visinelli@utah.edu}
\author{Paolo Gondolo}
\email{paolo@physics.utah.edu}
\affiliation{Department of Physics and Astronomy, University of Utah, 115 South 1400 East \#201, Salt Lake City, Utah 84112-0830, USA}

\date{\today}

\begin{abstract}
We study the parameter space of cold dark matter axions in two cosmological scenarios with non-standard thermal histories before Big Bang nucleosynthesis: the Low Temperature Reheating (LTR) cosmology and the kination cosmology. If the Peccei-Quinn symmetry breaks during inflation, we find more allowed parameter space in the LTR cosmology than in the standard cosmology  and less in the kination cosmology. On the contrary, if the Peccei-Quinn symmetry breaks after inflation, the Peccei-Quinn scale is orders of magnitude higher than standard in the LTR cosmology and lower in the kination cosmology. We show that the axion velocity dispersion may be used to distinguish some of these non-standard cosmologies. Thus, axion cold dark matter may be a good probe of the history of the Universe before Big Bang nucleosynthesis.
\end{abstract}


\maketitle

The standard cosmological model has been tested up to a temperature $T \sim 1\,{\rm MeV}$, or down to times as short as $\sim 1\,{\rm s}$, when Big Bang nucleosynthesis (BBN) occurred. The success of the BBN theory is due to its great precision in predicting the primordial abundance of light elements D, $^4$He and $^7$Li. For the success of BBN, the Universe must be radiation-dominated at temperatures $T \gtrsim 4 {\rm ~MeV}$ \cite{hannestad1}. However, due to lack of data prior to BBN, the history of the Universe in the pre-BBN epoch $T \gtrsim 4$ MeV is only indirectly inferred.

In the standard cosmology, radiation has been dominating the energy density of the Universe before BBN since the very early time at which inflation ended. How radiation was produced at the end of inflation from a state of negligible temperature is still a topic of active research: models include the decay of the inflaton field \cite{decay_inflaton, decay_inflaton1} and parametric resonance \cite{parametric_resonance}.

In alternative cosmological models, inflation may have ended at times close to BBN \cite{low_reheat_inflation}, or there might have been a period after inflation in which the dominant energy density was not in radiation but in some other exotic form, like the energy density of a scalar field \cite{turner_LTR}, or still more there could have been an injection of entropy into the radiation \cite{dine, steinhardt}.

A good probe of the history of the Universe before $\sim 1 {\rm s}$ is a relic particle that has survived from that period. Indeed, in the standard cosmology, dark matter relics like axions \cite{peccei, weinberg, wilczek} and Weakly Interacting Massive Particles (WIMPs) (see \cite{griest}) are produced when the Universe was $\sim 1 \mu{\rm s}$ or between $\sim 10 {\rm ns}$ and $\sim 10 {\rm ps}$ old (corresponding respectively to the age of the Universe at the QCD phase transition or at the WIMP freeze-out for WIMP masses between 10 GeV and 1 TeV). These dark matter relics are therefore excellent candidates to test the cosmological history of the Universe at very early epochs \cite{barrow}.

In this article we study cold dark matter (CDM) axions \cite{preskill, dine} as probes of the pre-BBN epoch, assuming that axions provide the totality of the CDM observed,
\begin{equation}
\Omega_a = \Omega_{\rm CDM},
\end{equation}
with \cite{komatsu}
\begin{equation} \label{CDM}
\Omega_{\rm CDM} h^2 = 0.1131 \pm 0.0034.
\end{equation}
Here $\Omega_a$ and $\Omega_{\rm CDM}$ are the axion and CDM energy densities in units of the critical density $\rho_{\rm crit}$, and $h$ is the Hubble constant in units of 100 km s$^{-1}$ Mpc$^{-1}$.

We analyze two non-standard pre-BBN cosmologies: the low-temperature reheating (LTR) cosmology \cite{dine, steinhardt, turner_LTR, lazarides, yamamoto, donald, moroi, kawasaki, chung, riotto, giudice, gelmini, gelmini1, drees, gelmini2} and the kination cosmology \cite{ford, kamionkowski_turner, spokoiny, joyce, tashiro, salati, profumo, rosati, pallis, chun, chung1, gomez}. In the LTR cosmology the expansion of the Universe after inflation is driven by a massive scalar field $\phi$ that decays and reheats the Universe. This stage lasts down to a temperature $T_{\rm RH}$, after which standard radiation-dominated cosmology applies. In the kination cosmology, the energy content of the Universe is dominated by the kinetic energy of a scalar field which evolves without entropy release down to a temperature $T_{\rm kin}$, after which standard cosmology begins.

We find that with respect to the standard cosmology, if the Peccei-Quinn symmetry in the axion theory breaks during inflation, the allowed axion parameter space is enlarged in the LTR cosmology and restricted in the kination cosmology. Instead, if the Peccei-Quinn symmetry breaks after inflation, the mass of cold dark matter axions  in the LTR cosmology is orders of magnitude smaller than its standard-cosmology value, and it is orders of magnitude greater in the kination cosmology.

Past work on axions in non-standard cosmologies has examined the parameter space of hot dark matter axions in the LTR and kination cosmologies \cite{kamionkowski}, and the cosmological bound $\Omega_a h^2 < 1$ for cold (i.e.\ non-thermal) axions in the LTR cosmology assuming that the Peccei-Quinn symmetry breaks after inflation \cite{dine, steinhardt, lazarides, yamamoto, kawasaki, riotto}. Our work studies axions as 100$\%$ of the cold dark matter, allows the PQ symmetry to break after or during inflation, and includes anharmonicities in the axion potential.

In Section \ref{production_of_axion_CDM}, we review the two most important mechanisms of axion production, the misalignment mechanism and axionic string decays, focusing on how to extend the standard results for these mechanisms to any non-standard cosmology. In Section \ref{Axion CDM in the standard scenario}, we review the present axion energy density $\Omega_a^{\rm std} h^2$ in the standard cosmology. In Section \ref{Axion CDM in the Low Temperature Reheating scenario}, we derive the density $\Omega_a^{\rm LTR} h^2$ and present our results for the axion parameter space for the LTR cosmology. In Section \ref{Axion CDM in the Kination scenario}, we do the same for the kination cosmology. In Section \ref{Discussion} we compare our results to previous work, discuss how our results may be modified by different choices of axionic string parameters, and suggest a possibility of distinguishing non-standard pre-BBN cosmologies observationally. Our conclusions are presented in Section \ref{Conclusions}.

\section{Production of axion dark matter} \label{production_of_axion_CDM}

The axion is a hypothetical pseudoscalar particle first introduced in the QCD sector of the Standard Model to solve the strong CP problem \cite{peccei, weinberg, wilczek}. Axion properties depend on several parameters (see e.g. \cite{kolb} and the recent reviews \cite{raffelt, kim_review}). They are: the scale $f_a$ at which the Peccei-Quinn (PQ) symmetry breaks, the PQ color anomaly $N$, the number of degenerate QCD vacua $N_d$, and the couplings of axions to quarks, leptons, and photons. Axion models include the KSVZ or hadronic model \cite{KSVZ}, in which there is no tree-level coupling between the axion and the standard model quarks and leptons, and the DFSZ model \cite{DFSZ}, in which the axion couples to standard model quarks and might couple to leptons. In both models, the axion-photon coupling is non-zero. Unless otherwise specified, our results apply to both KSVZ and DFSZ models (in both of which $N = N_d$).

Axions may account for the totality of the CDM observed \cite{preskill, dine, steinhardt, lazarides, yamamoto, kawasaki, riotto, turner_theta, davis, harari, turner, battye, khlopov, hagmann, shellard, wall_solution, string_simulation, fox, beltran, sikivie, hertzberg, bae, yang, hwang, baer, visinelli, hamann, wantz, mack}. This is accomplished for specific conditions on the parameters $f_a$ and $N$, on the expansion rate of the Universe $H_I$ at the end of inflation, and on the mechanism of axion production.

In this section, we review the two most important mechanisms to produce axion dark matter: vacuum realignment \cite{dine, preskill} and string decays \cite{davis, harari}. We focus on the modifications needed to go from the standard cosmology to non-standard cosmologies.

\subsection{Axions from vacuum realignment} \label{Axions from vacuum realignment}

We review the production of cold axions by the vacuum realignment mechanism, following the conventions in Ref.~\cite{visinelli}. The formulas in this section depend on the details of the cosmology before BBN only through the dependence of the Hubble expansion rate $H(T)$ on the temperature $T$. Different pre-BBN cosmologies differ in the choice of $H(T)$.

When the temperature of the Universe falls below $T \sim f_a$, the Peccei-Quinn symmetry breaks and the axion field $a(x)$ originates. The equation of motion for the misalignment angle $\theta(x) = a(x)/f_a$ is
\begin{equation}\label{eq_motion}
\ddot{\theta} + 3H(T)\,\dot{\theta} + \frac{1}{f_a^2}\,V'(\theta) = 0.
\end{equation}
Here a dot indicates a derivative with respect to time; $V'(\theta)$ is the derivative with respect to $\theta$ of the axion potential
\begin{equation}
V(\theta) = m_a^2(T)\,f_a^2\,(1-\cos\theta).
\end{equation}

The axion mass depends on temperature as \cite{gross}
\begin{equation} \label{axion_mass}
m_{a}(T) = \begin{cases}
1 & T \lesssim \Lambda,\\
b\,m_a \left(\frac{\Lambda}{T}\right)^4 & T \gtrsim \Lambda,\\
\end{cases}
\end{equation}
where we choose $N=1$, $\Lambda = 200$ MeV, $b=0.018$ \cite{beltran} and the zero-temperature axion mass \cite{weinberg}
\begin{equation}
\label{eq:axionmass}
m_a \equiv m_a(T=0) = 6.2\, {\rm \mu eV}\,f_{a,12}^{-1}.
\end{equation}
We use the notation $f_{a,12} = f_a / 10^{12}~{\rm GeV}$. For $T \gg \Lambda$, we can neglect $m_a^2(T)$ and so $V'(\theta)$ in Eq.~(\ref{eq_motion}). A solution to Eq.~(\ref{eq_motion}) in this case is then $\theta = \theta_i = {\rm const}$, where $\theta_i$ is the initial value of the misalignment angle. When $T \sim \Lambda$, coherent oscillations of the axion field begin. The temperature $T_1$ at which oscillations being is defined by the condition
\begin{equation} \label{T1}
3H(T_1) = m_a(T_1).
\end{equation}

The axion number density at $T_1$ can be written as \cite{turner, beltran, sikivie, hertzberg, visinelli}
\begin{equation} \label{numberdensity}
n_a(T_1) = \frac{1}{2}m_a(T_1)f_a^2\,\chi\,\langle\theta_i^2 f(\theta_i)\rangle.
\end{equation}
Here angle brackets indicate an average over the possible values of $\theta_i^2$ within a horizon volume, $\chi = 1.44$ is a fudge factor, and $f(\theta_i)$ is a corrective function accounting for anharmonicities in $V(\theta_i)$ \cite{turner}. (In our work on cold dark matter axions in standard cosmology \cite{visinelli}, we used  the symbol $\zeta$ for the parameter $\chi$, but here $\zeta$ is used for another quantity, see Section IIb.)

A complete study of the axion parameter space requires taking into account the function $f(\theta_i)$ in the whole range $-\pi < \theta_i < \pi$ \cite{visinelli, wantz}. The function $f(\theta_i)$ has been studied in various papers \cite{turner_theta, lyth, strobl, bae}. An analytic function interpolating between numerical results is \cite{visinelli}
\begin{equation} \label{anharmonicity_function}
f(\theta_i) = \left[\ln\left(\frac{e}{1-\theta_i^2/\pi^2}\right)\right]^{7/6}.
\end{equation}

The properties of the cosmological axion differ whether the axion field is present during inflation ($f_a>H_I/2\pi$) or not ($f_a<H_I/2\pi$).

For $f_a<H_I/2\pi$ (Scenario I), the initial misalignment angle $\theta_i$ acquires different values within the same Hubble horizon. Using Eq.~(\ref{anharmonicity_function}) we obtain $\langle\theta_i^2 f(\theta_i)\rangle = 8.77$ \cite{visinelli}, where the average is taken over the possible values of $\theta_i$.

If $f_a>H_I/2\pi$ (Scenario II), the PQ symmetry breaks during inflation and it is never restored. The initial misalignment angle $\theta_i$ takes only one value within a Hubble horizon. As noticed by Linde \cite{linde}, to a relatively small value of $\theta_i$ ($\sim 10^{-4}$) there corresponds a large  value of $f_a$ ($\sim 10^{16}{\rm ~GeV}$). This region of parameter space can thus accommodate values of $f_a$ of the order of the GUT scale ($\sim 10^{16}{\rm ~GeV}$) or above, however at the expense of tuning the initial misalignment angle to a small value which was the problem that the PQ mechanism tries to solve.

In Scenario II the mean $\langle\theta_i\rangle = \theta_i$ has a unique value within one Hubble volume, so
\begin{equation}
\langle\theta_i^2 f(\theta_i)\rangle = (\theta_i^2 + \sigma^2_{\theta})f(\theta_i).
\end{equation}
Here the variance
\begin{equation}
\sigma^2_{\theta} = \left(\frac{H_I}{2\pi\,f_a}\right)^2
\end{equation}
is due to the axion quantum fluctuations present at the end of inflation \cite{birrell}.

In Scenario II, axion isocurvature fluctuations are constrained by a combination of Cosmic Microwave Background (CMB), Baryon Acoustic Oscillations (BAO) and Supernovae (SN) measurements \cite{komatsu} as
\begin{equation} \label{isocurvature}
H_I < 4.168\times 10^{-5}\,\theta_i\,f_a.
\end{equation}
This constraint is obtained from an upper limit on the ratio of isocurvature and adiabatic power spectra at the wave number $k_0 = 0.002 ~{\rm Mpc^{-1}}$. Notice that this bound applies independently of the cosmological model between inflation and BBN because modes with wave number $k_0$ are still well outside the horizon at the epoch of recombination.

A further bound on $H_I$ comes from the non-detection of primordial gravitational waves. The combined CMB+BAO+SN measurements set an upper limit on $H_I$ given by \cite{komatsu}
\begin{equation} \label{gravitywavesbound}
H_I < 6.29 \times 10^{14}\,{\rm GeV}.
\end{equation}
In Scenario I ($f_a<H_I/2\pi$), this bound on $H_I$ implies an upper bound on $f_a$ equal to
\begin{equation}
f_a < 1.00\times 10^{14}\,{\rm GeV}, \quad \hbox{for $f_a<H_I/2\pi$}.
\end{equation}

Astrophysical considerations on the cooling time of white dwarfs yield the bound \cite{raffelt1}
\begin{equation} \label{whitedwarvesbound}
f_a > 4 \times 10^8\, {\rm GeV},
\end{equation}
valid for KSVZ axions. A similar bound from supernovae applies to other axion models \cite{raffelt1}.

The bounds in Eqs.~(\ref{gravitywavesbound}) and~(\ref{whitedwarvesbound}) apply independently of the pre-BBN history of the Universe.

\subsection{Axions from string decays} \label{Axions from string decays}
The spontaneous breaking of the Peccei-Quinn symmetry leads to the formation of topological defects such as axionic strings \cite{vilenkin}. In Scenario I ($f_a<H_I/2\pi$), the PQ symmetry breaks after inflation and cold axions produced by the decay of topological strings in the early Universe contribute a large fraction of the axion energy density. Instead, in Scenario II ($f_a>H_I/2\pi$), topological defects are washed out by inflation so axions from axionic strings are not present.

The present axion energy density produced by axionic strings $\rho_a^{\rm str}(T_0)$ is proportional to the present axion energy density produced by the misalignment mechanism $\rho_a^{\rm mis}(T_0)$ \cite{wall_solution, hagmann1, hagmann}. The ratio $\alpha$ between $\rho_a^{\rm str}(T_0)$ and $\rho_a^{\rm mis}(T_0)$ can be put in the form
\begin{equation} \label{proportionality}
\alpha \equiv \frac{\rho_a^{\rm str}(T_0)}{\rho_a^{\rm mis}(T_0)} = \frac{\xi \bar{r}N_d^2}{\zeta}.
\end{equation}
Here, following the notation in Ref.~\cite{hagmann} (however, we use $\zeta$ for their parameter $\chi$ to avoid confusion with our $\chi$ in Eq.~(\ref{numberdensity})): $N_d$ is the number of degenerate QCD vacua, $\bar{r}$ is the factor by which the axion comoving number density increases due to string decays, averaged over all possible processes that convert strings to axions, $\xi$ is a constant factor depending on the string network model, and $\zeta$ accounts for the uncertainties in the low-energy cutoff of the radiated axion field.

In the standard cosmology, the numerical values of these parameters have been discussed extensively both theoretically and via numerical simulations of string networks \cite{davis, harari, davis2, wall_solution, hagmann1, hagmann, sikivie_wall, battye, shellard, string_simulation, local_strings_results}. However, there is still disagreement about the numerical values of $\bar{r}$ and $\xi$ in the standard cosmology, $\bar{r}^{\rm std}$ and $\xi^{\rm std}$.

In the following we discuss each parameter in Eq.~(\ref{proportionality}) separately in standard cosmology and we extend previous theoretical results to obtain the values of the parameters in a generic non-standard cosmology.

{\it Parameter} $N_d. \quad$ In the standard cosmology, it is usually assumed that $N_d = 1$, because for $N_d > 1$ a domain wall problem may arise \cite{sikivie_wall}. In modified cosmologies we take the same value $N_d = 1$ as in the standard cosmology, because $N_d$ describes a property of the axion field.

{\it Parameter} $\bar{r}. \quad $ The value of $\bar{r}$ depends on the details of the axionic string relaxation toward lower energy configurations and on the energy spectrum of the radiated axions \cite{shellard}. In the standard cosmology it is (see Ref.~\cite{hagmann})
\begin{equation} \label{bar_r}
\bar{r}^{\rm std} = \begin{cases}
\ln(t_1/\delta), & \hbox{for a slow-oscillating string},\\
0.8, & \hbox{for a fast-oscillating string}.
\end{cases}
\end{equation}
Here, $t_1$ is the time at which the axion field starts to oscillate and $\delta$ is the string core size \cite{wall_solution}. In Eq.~(\ref{bar_r}), the first line corresponds to the string emission model in Refs.~\cite{davis, davis2, battye, shellard, string_simulation}, while the second line corresponds to the model in Refs.~\cite{wall_solution, harari, hagmann1, hagmann}.

The time $t_1$ can be expressed in terms of the corresponding temperature $T_1$ using the relation
\begin{equation} \label{time_temperature}
\frac{1}{2t} = H(T),
\end{equation}
valid in the standard cosmology. In the illustrative case $\delta = (10^{12} {\rm ~GeV})^{-1}$ and $T_1 = 1{\rm ~GeV}$, one finds $\bar{r}^{\rm std} \approx 70$, which is approximately the same value found in Refs.~\cite{davis, davis2, battye, shellard, string_simulation}.

To extend Eq.~(\ref{bar_r}) to non-standard cosmologies, we repeat its standard-cosmology derivation in Refs.~\cite{hagmann, harari} but change the relation between time and Hubble parameter in Eq.~(\ref{time_temperature}) to that appropriate for a non-standard cosmology. We consider a generic dependence of the scale factor $a(t)$ on time $t$,
\begin{equation} \label{a(t)}
a(t) \propto t^{\beta},
\end{equation}
where $\beta$ is a constant that depends on the details of the modified cosmology. For example, $\beta = 2/3$ for the matter-dominated Universe and the LTR cosmology, $\beta = 1/2$ for the radiation-dominated Universe, and $\beta = 1/3$ for the kination cosmology. Eq.~(\ref{time_temperature}) then becomes
\begin{equation}
H(t) \equiv \frac{\dot{a}(t)}{a(t)} = \frac{\beta}{t}.
\end{equation}

Harari and Sikivie \cite{harari} derive the axion number density from string decays $n_a^{\rm str}(t)$ from the equations
\begin{equation} \label{dn/dt}
\frac{d n_a^{\rm str}(t)}{dt} = \frac{1}{\omega(t)}\frac{d\rho_r}{dt} - 3H(t)n_a^{\rm str}(t),
\end{equation}
and
\begin{equation} \label{drho/dt}
\frac{d\rho_r}{dt} = -\frac{d\rho_s}{dt} - 2H(t)\rho_s.
\end{equation}
Here, $\rho_r$ is the energy density of the radiated axions, $\omega(t)$ is the average energy of axions radiated in string decay processes at time $t$ \cite{hagmann}, and $\rho_s$ is the energy density in strings, given by
\begin{equation} \label{rho_s}
\rho_s = \frac{\xi\,N_d^2}{\zeta}\,\frac{\pi\,f_a^2\ln(t/\delta)}{t^2}.
\end{equation}
From Eqs.~(\ref{dn/dt}),~(\ref{drho/dt}) and~(\ref{rho_s}) we obtain
\begin{equation}
n_a^{\rm str}(t_1) = \frac{\xi\,N_d^2\,f_a^2}{\zeta}\,\frac{2\pi\,(1-\beta)}{t_1^{3\beta}}\int_{t_{PQ}}^{t_1} \frac{dt}{t^{3-3\beta}}\frac{\ln(t/\delta)}{\omega(t)},
\end{equation}
where $t_{PQ} \ll t_1$ is the time at which the PQ phase transition occurs. The formula to obtain the parameter $\bar{r}$ follows from Eq.~(2.13) in Ref.~\cite{hagmann},
\begin{equation}
n_a^{\rm str}(t) = \frac{\xi\,\bar{r}N_d^2\,f_a^2}{\zeta\,t},
\end{equation}
which, evaluated at time $t_1$, gives
\begin{equation} \label{bar_r1}
\bar{r} = \frac{2\pi\,(1-\beta)}{t_1^{3\beta-1}}\int_{t_{PQ}}^{t_1} \frac{dt}{t^{3-3\beta}}\frac{\ln(t/\delta)}{\omega(t)}.
\end{equation}

The function $\omega(t)$ depends on the model for the energy spectrum of the emitted axions. For slow-oscillating strings, Davis \cite{davis} argues that the energy spectrum of the axions radiated at time $t$ is peaked around $2\pi/t$, and finds $\omega(t) = 2\pi/t$. Using this expression of $\omega(t)$ in Eq.~(\ref{bar_r1}) gives
\begin{equation} \label{bar_r_sharp}
\bar{r} = \begin{cases}
\frac{1-\beta}{3\beta-1}\ln(t_1/\delta), & \hbox{for $\beta \neq 1/3$},\\
\frac{2}{3}\ln(t_1/t_{\rm PQ})\ln(t_1/\delta), & \hbox{for $\beta = 1/3$}.
\end{cases}
\end{equation}
In particular, for the standard cosmology $\beta = 1/2$ and $\bar{r}^{\rm std} = \ln(t_1/\delta)$, as in the first line of Eq.~(\ref{bar_r}).

For fast-oscillating strings, Harari and Sikivie \cite{harari} argue that the energy spectrum of the radiated axions is broad, with a low-energy cutoff at energy $\pi/t_1$ and a high-energy cutoff at energy $\pi/\delta$. They find $\omega(t) = (2\pi/t)\ln(t/\delta)$. Eq.~(\ref{bar_r1}) with this expression of $\omega(t)$ leads to $\bar{r}^{\rm std} = 1$ for the standard cosmology. Numerical simulations \cite{wall_solution, hagmann1, hagmann} favor a slightly smaller value of $\bar{r}^{\rm std}$, namely 0.8 as quoted in Eq.~(\ref{bar_r}). Therefore, we decided to multiply Eq.~(\ref{bar_r1}) by 0.8. Hence, for the fast-oscillating strings,
\begin{equation} \label{bar_r_broad}
\bar{r} = \begin{cases}
0.8\,\frac{1-\beta}{3\beta-1}, & \hbox{for $\beta \neq 1/3$},\\
0.8\,\frac{2}{3}\,\ln(t_1/t_{\rm PQ}), & \hbox{for $\beta = 1/3$}.
\end{cases}
\end{equation}
In the particular case of the LTR cosmology ($\beta = 2/3$), Eq~(\ref{bar_r_broad}) gives $\bar{r}^{\rm LTR} = 0.27$, while in the kination cosmology ($\beta = 1/3$), Eq.~(\ref{bar_r_broad}) gives $\bar{r}^{\rm kin} = 0.53\,\ln(t_1/t_{\rm PQ})$.

In presenting our results, we use the value of $\bar{r}$ for fast-oscillating strings, Eq.~(\ref{bar_r_broad}). We discuss the alternative choice of Eq.~(\ref{bar_r_sharp}) in Section~\ref{Discussion}.

{\it Parameter} $\xi. \quad $ The value of $\xi^{\rm std}$ in the standard cosmology has been discussed in the literature and different authors quote different results \cite{hagmann1, hagmann, string_simulation,local_strings_results}. Numerical simulations for an evolving string network in Ref.~\cite{local_strings_results} yield $\xi^{\rm std} \sim 13$, while simulations in Refs.~\cite{hagmann1, hagmann, string_simulation} give $\xi^{\rm std} \sim 1$.

The value of $\xi$ changes in a modified cosmological scenario. The authors in Ref.~\cite{string_matter} outline a method to estimate $\xi$ in a modified cosmology in which the Universe is matter-dominated from the value $\xi^{\rm std}$ for a radiation-dominated Universe. Here we generalize the results of Ref.~\cite{string_matter} to a generic cosmology with arbitrary $\beta$, following their method. We define the characteristic length $L$ for an axionic string of energy density $\rho$ and tension $\mu$ per unit length through $\rho = \mu/L^2$.  The parameter $\xi$ appears in the time dependence of the string length $L$ as $L = t/\sqrt{\xi}$. The method of Ref.~\cite{string_matter} consists in computing $\xi$ as
\begin{equation}
\label{eq:xidef}
\xi = (\gamma_0\,H\,t)^2,
\end{equation}
where $\gamma_0$ is the fixed-point value of Eq.~(14) in Ref.~\cite{string_matter},
\begin{equation} \label{yamaguchi}
\frac{d\gamma}{dt} = -\frac{H}{2}\left\{c\gamma^2 + [2\frac{\dot{H}}{H^2} + 3]\gamma - 1\right\}.
\end{equation}
Here $\gamma = (H(t) L)^{-1}$, and $c>0$ is a constant determined by the value of $\xi$ for a radiation-dominated universe.

For a generic $\beta$, we have
\begin{equation}
\frac{\dot{H}}{H^2} = -\frac{1}{\beta},
\end{equation}
and the fixed-point of Eq.~(\ref{yamaguchi}) follows from setting its right-hand side to zero,
\begin{equation} \label{gamma_0}
\gamma_0 = \frac{2-3\beta + \sqrt{(4c+9)\beta^2 -12\beta+4}}{2\beta \,c}.
\end{equation}
Then from Eq.~(\ref{eq:xidef}) and $H=\beta/t$, we find
\begin{equation} \label{xi1}
\xi = \frac{1}{4c^2}\left(2-3\beta + \sqrt{(4c+9)\beta^2 -12\beta+4}\right)^2.
\end{equation}
The constant $c$ is fixed from the requirement that $\xi = \xi^{\rm std}$ in a radiation-dominated Universe, $\beta=1/2$. This gives
\begin{equation}\label{c}
c = \frac{1+2\sqrt{\xi^{\rm std}}}{4\xi^{\rm std}}.
\end{equation}
This is to be substituted in Eq.~(\ref{xi1}) to find the value of $\xi$ in the non-standard cosmology.

For $\xi^{\rm std} = 1$, we find $c = 3/4$ and
\begin{equation} \label{xi}
\xi = \frac{4}{9}\left(2-3\beta + 2\sqrt{3\beta^2 -3\beta+1}\right)^2\quad\hbox{(for $\xi^{\rm std} = 1$)}.
\end{equation}

Then, for the LTR cosmology, in which $\beta=2/3$, we find $\xi^{\rm LTR} = 16/27 = 0.5926$; for the kination cosmology, in which $\beta = 1/3$, we find $\xi^{\rm kin} = 4(7+4\sqrt{3})/27 = 2.0634$.

The choice $\xi^{\rm std} = 13$ is discussed in Section \ref{Discussion}.

{\it Parameter} $\zeta. \quad$ In the standard cosmology, $\zeta \sim 1$ \cite{hagmann}. To find $\zeta$ in a generic cosmology we use the fact that on dimensional grounds, $\zeta$ is of order $\sqrt{\xi}$ \cite{hagmann}, so that to a change $\Delta \xi$ there corresponds a change $\Delta \zeta/2$. However, the theoretical uncertainty on $\zeta$ is around 50$\%$ \cite{hagmann}, higher than the difference $\Delta \xi$ due to the change in the cosmology used. We thus consider $\zeta$ constant and equal in all cosmological scenarios.

\section{Axion CDM in the standard cosmology} \label{Axion CDM in the standard scenario}

In this section we review the derivation of the present cold axions density, assuming that the Universe follows the standard cosmology. Coherent oscillations begin in a radiation-dominated Universe, for which the Hubble parameter is
\begin{equation} \label{hubble_std}
H(T) = \sqrt{\frac{8\pi^3}{90}g_*(T)}\frac{T^2}{M_{Pl}} \simeq 1.66\sqrt{g_*(T)}\frac{T^2}{M_{Pl}}.
\end{equation}
Here $M_{Pl} \sim 1.22\times 10^{19} {\rm ~GeV}$ is the Planck mass. We approximate the relativistic degrees of freedom $g_*(T)$ in the range of temperature we are interested in by
\begin{equation} \label{g*}
g_*(T) =
\begin{cases}
61.75, & \hbox{for $T \gtrsim \Lambda$},\\
10.75, & \hbox{for $\Lambda \gtrsim T \gtrsim 4 {\rm ~MeV}$},\\
 3.36, & \hbox{for $T \lesssim 4 {\rm ~MeV}$}.
\end{cases}
\end{equation}
Eqs.~(\ref{axion_mass}),~(\ref{T1}) and~(\ref{hubble_std}) provide the temperature $T_1^{\rm std}$ at which the axion field begins to oscillate:
\begin{equation} \label{t1std}
T_1^{\rm std} = \begin{cases}
123{\rm ~GeV}\,g_*^{-1/4}(T_1^{\rm std})f_{a,12}^{-1/2}, & \hbox{for $T_1^{\rm std} \lesssim \Lambda$},\\
871{\rm ~MeV}\,g_*^{-1/12}(T_1^{\rm std})f_{a,12}^{-1/6}, & \hbox{for $T_1^{\rm std} \gtrsim \Lambda$}.
\end{cases}
\end{equation}
For the entropy degrees of freedom $g_{*S}(T)$ we use the approximation
\begin{equation}
g_S(T) =
\begin{cases}
g_*(T), & \hbox{for $T \gtrsim 4 {\rm ~MeV}$},\\
3.91, & \hbox{for $T \lesssim 4 {\rm ~MeV}$}.
\end{cases}
\end{equation}
Entropy conservation in the standard cosmology leads to the following relation between the scale factor $a^{\rm std}(T)$ and the temperature $T$:
\begin{equation} \label{scale_factor}
g_{*S}^{1/3}(T)\,T\,a^{\rm std}(T) = {\rm constant}.
\end{equation}

The present axion number density $n_a^{\rm std}(T_0)$ is computed from the number density at $T_1^{\rm std}$, Eq.~(\ref{numberdensity}), assuming conservation of entropy and of the number of axions in a comoving volume,
\begin{equation}
\label{eq:nastd}
n_a^{\rm std}(T_0) =
n_a(T_1^{\rm std})\,\left[\frac{a^{\rm std}(T_1^{\rm std})}{a^{\rm std}(T_0)}\right]^3.
\end{equation}

The misalignment mechanism gives a contribution to the present axion energy density  $\rho^{\rm std,mis}_a = m_a\,n_a^{\rm std}(T_0)$. In units of the critical density $\rho_{\rm crit}$ we have
\begin{equation} \label{standard_density}
\Omega^{\rm std,mis}_a = \frac{m_a n_a^{\rm std}(T_1^{\rm std})}{\rho_{\rm crit}}\frac{g_{*S}(T_0)}{g_{*S}(T^{\rm std}_1)}\left(\frac{T_0}{T^{\rm std}_1}\right)^3.
\end{equation}
Inserting the numerical values we obtain
\begin{equation} \label{standard_density_numerical}
\Omega^{\rm std,mis}_a h^2 = \begin{cases}
1.32\,g_*^{-5/12}(T_1^{\rm std})\,\langle\theta_i^2\,f(\theta_i)\rangle\,f_{a,12}^{7/6}, & \hbox{for $f_a < \hat{f}_a$},\\
9.23\times 10^{-3}\,g_*^{-1/4}(T_1^{\rm std})\,\langle\theta_i^2\,f(\theta_i)\rangle\,f_{a,12}^{3/2}, & \hbox{for $f_a > \hat{f}_a$}.
\end{cases}
\end{equation}
Here $h$ is the Hubble constant in units of 100 km ${\rm s}^{-1} {\rm Mpc}^{-1}$, and $\hat{f}_a$ is the PQ scale at which the two expressions in Eq.~(\ref{standard_density_numerical}) are equal. For the values for $g_*(T)$ as in Eq.~(\ref{g*}),
\begin{equation} \label{PQ_scale}
\hat{f}_a = 9.9\times 10^{16}{\rm ~GeV}.
\end{equation}

String decays give a contribution to the present axion energy density
\begin{equation} \label{strings}
\Omega_a^{\rm std,str} = \alpha^{\rm std}\,\Omega_a^{\rm std,mis} = 0.164 \,\Omega_a^{\rm std,mis},
\end{equation}
where for $\alpha^{\rm std}$ in Eq.~(\ref{proportionality}) we have taken $N_d = 1$, $\xi^{\rm std} = 1$, $\bar{r}^{\rm std} = 0.8$ \cite{hagmann} and $\zeta = 4.9$, consistently with our previous work \cite{visinelli}. In Section VI we discuss the modifications to Eq.~(\ref{strings}) and to our results if the values $\bar{r}^{\rm std} = 70$, $\xi^{\rm std} =13$ \cite{davis, davis2, battye, shellard, string_simulation, local_strings_results} are used.

The present axion energy density in the standard cosmology is then given by the sum of the misalignment mechanism and string decays contributions
\begin{equation} \label{energy_density_total_std}
\Omega^{\rm std}_a = \Omega^{\rm std,mis}_a + \Omega^{\rm std,str}_a .
\end{equation}

If the PQ symmetry breaks after the end of inflation (Scenario I, $f_a<H_I/2\pi$), there is only one PQ scale $f_a$ for which the totality of cold dark matter is made of axions. There correspondingly are also a single value for the axion mass $m_a$ from Eq.~(\ref{eq:axionmass}), and for the temperature $T_1^{\rm std}$, from Eq.~(\ref{t1std}). Using the observed value of $\Omega_{\rm CDM} h^2$ in Eq.~(\ref{CDM}) and the expressions for $\Omega_a^{\rm std}$ in Section II with $\langle\theta_i^2\,f(\theta_i)\rangle = 8.77$, we have \cite{visinelli}
\begin{equation} \label{naturalscale}
f_a^{\rm std} = (7.27 \pm 0.25)\times 10^{10}\,{\rm GeV},
\end{equation}
\begin{equation}
m_a^{\rm std} = 85 \pm 3\, {\rm \mu eV},
\end{equation}
and
\begin{equation} \label{T1natural}
T_1^{\rm std} = 956{\rm MeV}.
\end{equation}

In Scenario II no axions from axionic strings are present. The parameter space is bounded by the non-detection of axion isocurvature fluctuations in the CMB spectrum, see Eq.~(\ref{isocurvature}). One parameter in Eq.~(\ref{isocurvature}) can be eliminated by using the equality between the axion energy density and the CDM energy density. In the standard cosmology, the isocurvature bound is
\begin{equation} \label{leftmostboundary}
H_{I,12} < \begin{cases}
2.89\times 10^{-5}\,\sqrt{f(\theta_i)}\,f_{a,12}^{5/12}, & \hbox{for $f_a < 9.9\times 10^{16}{\rm ~GeV}$},\\
1.96\times 10^{-4}\,f_{a,12}^{1/4}, & \hbox{for $9.9\times 10^{16}{\rm ~GeV} < f_a$}.
\end{cases}
\end{equation}
It can be approximated by
\begin{equation} \label{leftmostboundary1}
H_{I,12} < \begin{cases}
1.31\times 10^{-4}\,f_{a,12}, & \hbox{for $f_a < 7.6 \times 10^{10} {\rm ~GeV}$},\\
2.89\times 10^{-5}\,f_{a,12}^{5/12}, & \hbox{for $7.6 \times 10^{10} {\rm ~GeV} < f_a < 9.9\times 10^{16} {\rm ~GeV}$},\\
1.96\times 10^{-4}\,f_{a,12}^{1/4}, & \hbox{for $9.9\times 10^{16} {\rm ~GeV} < f_a$},
\end{cases}
\end{equation}
where $H_{I,12} = H_I/10^{12}\,{\rm GeV}$ and $f_{a,12} = f_a/10^{12}{\rm ~GeV}$. The power-law dependence of the bound on $f_{a,12}$ changes twice:  at $f_a \sim 10^{11}{\rm ~GeV}$ due to the effects of anharmonicities \cite{visinelli}, and at $f_a = 9.9\times 10^{16}{\rm ~GeV}$, due to the change in the dependence of the axion mass with temperature.

\section{Axion CDM in the Low Temperature Reheating cosmology} \label{Axion CDM in the Low Temperature Reheating scenario}

In the low-temperature reheating (LTR) cosmology \cite{dine, steinhardt, turner_LTR, lazarides, yamamoto, donald, moroi, kawasaki, chung, riotto, giudice, gelmini, gelmini1, drees, gelmini2}, the Universe after inflation is dominated by a massive decaying scalar field $\phi$ down to the reheating temperature $T_{\rm RH}$. The reheating temperature is defined as the temperature at which the decay width $\Gamma_{\phi}$ of the scalar field $\phi$ is equal to the Hubble expansion rate $H(T)$ \cite{moroi, kawasaki, chung, riotto, giudice, gelmini, gelmini1, drees, gelmini2},
\begin{equation}
T_{\rm RH} \equiv \left(\frac{90}{8\pi^3\,g_*(T_{\rm RH})}\right)^{1/4}\sqrt{\Gamma_{\phi}\,M_{Pl}}.
\end{equation}

At $T>T_{\rm RH} $, the Universe follows the LTR cosmology; at $T=T_{\rm RH}$, it transitions to the usual radiation-dominated era. At $T<T_{\rm RH}$, the Hubble expansion rate is given by Eq.~(\ref{hubble_std}); at $T>T_{\rm RH}$, the Hubble rate $H(T)$ depends on the scale factor $a^{\rm LTR}(t)$ as in a matter-dominated epoch \cite{turner_LTR, lazarides, yamamoto, donald, moroi, kawasaki, chung, riotto},
\begin{equation}
H(T) = H(T_{\rm RH})\left(\frac{a^{\rm LTR}(T_{\rm RH})}{a^{\rm LTR}(T)}\right)^{3/2} =  \sqrt{\frac{8\pi^3}{90}g_*(T_{\rm RH})}\frac{T_{\rm RH}^2}{M_{Pl}}\,\left(\frac{a^{\rm LTR}(T_{\rm RH})}{a^{\rm LTR}(T)}\right)^{3/2},
\quad \hbox{for $T>T_{\rm RH}$.} \end{equation}
Here $H(T_{\rm RH})$ is given by its expression in the radiation-dominated epoch.

Using the relation between the scale factor $a^{\rm LTR}(T)$ and the temperature $T$ during the LTR epoch \cite{lazarides},
\begin{equation} \label{scale_factor_LTR}
g_*(T)^{2/3}\,T^{8/3}\,a^{\rm LTR}(T) = {\rm constant}, \quad \hbox{for $T>T_{\rm RH}$},
\end{equation}
and the usual relation during the radiation-dominated epoch,
\begin{equation}
g_*(T)^{1/3}\,T\,a^{\rm std}(T) = {\rm constant},\quad \hbox{for $T<T_{\rm RH}$},
\end{equation}
we find the Hubble expansion rate
\begin{equation} \label{H_nonstandard}
H(T) = \begin{cases}
\sqrt{\frac{8\pi^3}{90}g_*(T)}\,\frac{T^2}{M_{Pl}}, & \hbox{for $T < T_{\rm RH}$},\\
\sqrt{\frac{8\pi^3}{90}g_{RH}(T)}\,\frac{T^4}{T^2_{RH}M_{Pl}}, & \hbox{for $T > T_{\rm RH}$},\\
\end{cases}
\end{equation}
where
\begin{equation}
g_{RH}(T) = \frac{g_*^2(T)}{g_*(T_{\rm RH})}.
\end{equation}
The relation in Eq.~(\ref{scale_factor_LTR}) shows that the evolution of the Universe during the LTR stage is non-adiabatic.

In the standard cosmology, the axion field starts to oscillate at a temperature $T_1^{\rm std}$ given by Eq.~(\ref{T1}) with the standard expansion rate $H(T)$ in the right-hand side. In the LTR cosmology, $H(T)$ differs from the standard expression at $T>T_{\rm RH}$, and the axion field may start oscillating at a different temperature $T_1^{\rm LTR}$.

More precisely, if the standard temperature $T_1^{\rm std}$ is less than $T_{\rm RH}$, then the axion field starts to oscillate when the Universe is radiation-dominated. Moreover, since $H(T)$ is the same in both cosmologies at $T<T_{\rm RH}$, the oscillations start at the temperature $T_1^{\rm LTR}=T_1^{\rm std}$ given in Eq.~(\ref{t1std}). In this case the results of Section \ref{Axion CDM in the standard scenario} apply.

On the other hand, if $T_1^{\rm std}$ would be larger than $T_{\rm RH}$, then the LTR temperature $T_1^{\rm LTR}$ will be smaller than $T_1^{\rm std}$. In this case, the axion field starts to oscillate when the Universe is dominated by the decay of the massive scalar field $\phi$. The temperature $T_1^{\rm LTR}$ follows from Eq.~(\ref{T1}) with $H(T)$ given by the first line of Eq.~(\ref{H_nonstandard}). Since the dependence of $H(T)$ on $T$ steepens from $T^2$ to $T^4$ as $T$ becomes greater than $T_{\rm RH}$, it follows that
\begin{equation} \label{T_1-relation-LTR}
T_1^{\rm LTR} < T_1^{\rm std}.
\end{equation}
With $T_{\rm RH,MeV} = T_{\rm RH}/{\rm MeV}$, we find
\begin{equation}
T_1^{\rm LTR} = \begin{cases}
\left(m_a M_{Pl} T_{\rm RH}^2 \sqrt{\frac{5}{4\pi^3\,g_{RH}(T_1^{\rm LTR})}}\right)^{1/4}, & \hbox{for $T_1^{\rm LTR} \lesssim \Lambda$},\\
\left(b m_a M_{Pl} T_{\rm RH}^2 \Lambda^{4}\sqrt{\frac{5}{4\pi^3\,g_{RH}(T_1^{\rm LTR})}}\right)^{1/8}, & \hbox{for $T_1^{\rm LTR} \gtrsim \Lambda$}.
\end{cases}
\end{equation}
Numerically,
\begin{equation}\label{T1LTR}
T_1^{\rm LTR} = \begin{cases}
351{\rm ~MeV}\,g_{RH}^{-1/8}(T_1^{\rm LTR})\,T_{\rm RH,MeV}^{1/2}\,f_{a,12}^{-1/4}, & \hbox{for $T_1^{\rm LTR} \lesssim \Lambda$},\\
160{\rm ~MeV}\,g_{RH}^{-1/16}(T_1^{\rm LTR})\,T_{\rm RH,MeV}^{1/4}\,f_{a,12}^{-1/8}, & \hbox{for $T_1^{\rm LTR} \gtrsim \Lambda$}.
\end{cases}
\end{equation}
To summarize, if $T_{\rm RH} < \Lambda$,
\begin{equation}
T_1^{\rm LTR} = \begin{cases}
123{\rm ~GeV}\,g_*^{-1/4}(T_1^{\rm std})f_{a,12}^{-1/2}, & \hbox{for $T_1^{\rm std} < T_{\rm RH}$},\\
351{\rm ~MeV}\,g_{RH}^{-1/8}(T_1^{\rm LTR})\,T_{\rm RH,MeV}^{1/2}\,f_{a,12}^{-1/4}, & \hbox{for $T_{\rm RH} < T_1^{\rm LTR} \lesssim \Lambda$},\\
160{\rm ~MeV}\,g_{RH}^{-1/16}(T_1^{\rm LTR})\,T_{\rm RH,MeV}^{1/4}\,f_{a,12}^{-1/8}, & \hbox{for $\Lambda \lesssim T_1^{\rm LTR}$};
\end{cases}
\end{equation}
if $T_{\rm RH} > \Lambda$,
\begin{equation}
T_1^{\rm LTR} = \begin{cases}
123{\rm ~GeV}\,g_*^{-1/4}(T_1^{\rm std})f_{a,12}^{-1/2}, & \hbox{for $T_1^{\rm std} \lesssim \Lambda$}, \\
871{\rm ~MeV}\,g_*^{-1/12}(T_1^{\rm std})f_{a,12}^{-1/6}, & \hbox{for $\Lambda \lesssim T_1^{\rm std}  < T_{\rm RH}$}, \\
160{\rm ~MeV}\,g_{RH}^{-1/16}(T_1^{\rm LTR})\,T_{\rm RH,MeV}^{1/4}\,f_{a,12}^{-1/8}, & \hbox{for $T_{\rm RH} < T_1^{\rm std}$} .
\end{cases}
\end{equation}

Also the present axion energy density is modified from the standard case if $T_1^{\rm std} > T_{\rm RH}$.  We examine the misalignment mechanism and the production in string decays separately.

String decays give a contribution to the present axion energy density
\begin{equation} \label{strings_LTR}
\Omega_a^{\rm LTR,str} = \alpha^{\rm LTR}\,\Omega_a^{\rm LTR,mis} = 0.032\,\Omega_a^{\rm LTR,mis},
\end{equation}
where we used the values $N_d = 1$, $\bar{r}^{\rm LTR} = 0.27$, $\xi^{\rm LTR} = 16/27$ and $\zeta = 4.9$ in Eq.~(\ref{proportionality}), consistently with the discussion in Section \ref{Axions from string decays}.

For the misalignment mechanism, the axion number density at the present time can be found from Eq.~(\ref{numberdensity}) with $T_1 = T_1^{\rm LTR}$ using the conservation of axion number in a comoving volume, $n_a(T) \propto a^{-3}(T)$. This gives
\begin{equation} \label{n(RH)}
n_a^{\rm LTR}(T_0) =
\begin{cases}
n_a(T_1^{\rm std})\,\left(\frac{a^{\rm std}(T_1^{\rm std})}{a^{\rm std}(T_0)}\right)^3 , & \hbox{for $T_1^{\rm std} < T_{\rm RH}$},\\
n_a(T_1^{\rm LTR})\,\left(\frac{a^{\rm LTR}(T_1^{\rm LTR})}{a^{\rm std}(T_0)}\right)^3 , & \hbox{for $T_1^{\rm std} > T_{\rm RH}$}.
\end{cases}
\end{equation}
Here $n_a(T_1)$ is the function given in Eq.~(\ref{numberdensity}). One clearly has
\begin{equation}
n_a^{\rm LTR}(T_0) = n_a^{\rm std}(T_0) \quad \hbox{for $T_1^{\rm std} < T_{\rm RH}$.}
\end{equation}

For $T_1^{\rm std} > T_{\rm RH}$, one obtains a different axion density. To understand the origin of the difference, it is convenient to introduce the ratio between the present density $n_a^{\rm LTR}(T_0)$ in the LTR cosmology, and the present density $n_a^{\rm std}(T_0)$ in Eq.~(\ref{eq:nastd}) that would ensue if the cosmology were standard at temperatures $T>T_{\rm RH}$.
We write, for $T_1^{\rm std} > T_{\rm RH}$,
\begin{equation}
\frac{n_a^{\rm LTR}(T_0)}{n_a^{\rm std}(T_0)} =
\frac{N^{\rm LTR}}{N^{\rm std}}\,\frac{V^{\rm LTR}}{V^{\rm std}} ,
\end{equation}
where
\begin{equation} \label{N/N_std}
\frac{N^{\rm LTR}}{N^{\rm std}} = \frac{n_a(T_1^{\rm LTR})}{n_a(T_1^{\rm std})} \left( \frac{a^{\rm std}(T_1^{\rm LTR})}{a^{\rm std}(T_1^{\rm std})} \right)^3
\end{equation}
is the standard-cosmology ratio of the comoving number of axions $N^{\rm LTR}$  at the temperature $T_1^{\rm LTR}$ to the comoving number of axions $N^{\rm std}$ at the temperature $T_1^{\rm std}$,
and
\begin{equation} \label{V/V_std}
\frac{V^{\rm LTR}}{V^{\rm std}} = \left(\frac{a^{\rm LTR}(T_1^{\rm LTR})}{a^{\rm std}(T_1^{\rm LTR})}\right)^3 ,
\end{equation}
is the ratio of the LTR-cosmology volume $V^{\rm LTR}$ to the standard-cosmology volume $V^{\rm std}$ at the temperature $T_1^{\rm LTR}$.

The ratio $N^{\rm LTR}/N^{\rm std}$ accounts for the fact that coherent oscillations in the axion field start at a different temperature in the LTR cosmology compared to the standard cosmology. The ratio $V^{\rm LTR}/V^{\rm std}$ accounts for the fact that at temperature $T_1^{\rm LTR}$ the scale factors, and so the volumes, in the LTR and in the standard cosmologies differ due to entropy production from the decay of the scalar field in the LTR case.

Using the relations between temperature and scale factor during the radiation and LTR epochs, Eqs.~(\ref{scale_factor}) and~(\ref{scale_factor_LTR}) respectively, we find
\begin{equation} \label{N/N_std1}
\frac{N^{\rm LTR}}{N^{\rm std}} = \frac{g_{*S}(T_1^{\rm std})}{g_{*S}(T_1^{\rm LTR})}\left(\frac{T_1^{\rm std}}{T_1^{\rm LTR}}\right)^7,
\end{equation}
and
\begin{equation} \label{V/V_std1}
\frac{V^{\rm LTR}}{V^{\rm std}} = \frac{g_{*S}(T_1^{\rm LTR})}{g_{*S}(T_{\rm RH})}\frac{g^2_*(T_{\rm RH})}{g^2_*(T_1^{\rm LTR})}\,\left(\frac{T_{\rm RH}}{T_1^{\rm LTR}}\right)^5.
\end{equation}

From $T_1^{\rm LTR} < T_1^{\rm std}$ (see Eq.~(\ref{T_1-relation-LTR})), we find that $N^{\rm LTR} > N^{\rm std}$. But for $T_1^{\rm LTR} > T_{\rm RH}$, $V^{\rm LTR} < V^{\rm std}$. The latter factor dominates, and $n_a^{\rm LTR}(T_0)$ is less than $n_a^{\rm std}(T_0)$.

The present axion energy density from the misalignment mechanism follows as, in units of the critical density,
\begin{equation} \label{compare}
\Omega_a^{\rm LTR,mis} = \frac{m_a n_a^{\rm LTR}(T_0)}{\rho_{\rm crit}} = \begin{cases}
\Omega_a^{\rm std,mis}, & \hbox{for $T_1^{\rm std} < T_{\rm RH}$},\\
\Omega_a^{\rm std,mis} \frac{N^{\rm LTR}}{N^{\rm std}}\,\frac{V^{\rm LTR}}{V^{\rm std}}, & \hbox{for $T_1^{\rm std} > T_{\rm RH}$}.
\end{cases}
\end{equation}
Here $\Omega_a^{\rm std,mis}$ is given in Eq.~(\ref{standard_density}).

The first line of Eq.~(\ref{compare}), valid for $T_1^{\rm std} < T_{\rm RH}$, is numerically equal to Eq.~(\ref{standard_density_numerical}). The second line of Eq.~(\ref{compare}), valid for $T_1^{\rm std} > T_{\rm RH}$, is
\begin{equation} \label{Omega_LTR_numerical}
\Omega_a^{\rm LTR,mis} h^2 = \begin{cases}
1.78\times 10^{-6}\,\langle\theta_i^2\,f(\theta_i)\rangle\,g_{RH}^{-1/4}(T_1^{\rm LTR})\,f_{a,12}^{3/2}\,T_{\rm RH,MeV}^2, & {\rm for}\, f_a < \hat{f}_a(T_{\rm RH}),\\
7.46\times 10^{-8}\,\langle\theta_i^2\,f(\theta_i)\rangle\,f_{a,12}^2\,T_{\rm RH,MeV}, & {\rm for}\, f_a > \hat{f}_a(T_{\rm RH}),
\end{cases}
\end{equation}
where
\begin{equation} \label{PQ_scale_LTR}
\hat{f}_a(T_{\rm RH}) = 5.69\times 10^{14}{\rm ~GeV}\,g_{RH}^{-1/2}(T_1^{\rm LTR})\,T_{\rm RH,MeV}^2
\end{equation}
is the PQ scale at which the two lines in Eq.~(\ref{Omega_LTR_numerical}) match.

The present axion energy density in the LTR cosmology is given by the sum of the misalignment mechanism and the string decay contributions
\begin{equation} \label{energy_density_total_LTR}
\Omega^{\rm LTR}_a = \Omega^{\rm LTR,mis}_a + \Omega^{\rm LTR,str}_a = \begin{cases}
\Omega_a^{\rm std,mis}\,(1+\alpha^{\rm std}), & \hbox{for $T_1^{\rm std} < T_{\rm RH}$},\\
\Omega_a^{\rm std,mis} \frac{N^{\rm LTR}}{N^{\rm std}}\,\frac{V^{\rm LTR}}{V^{\rm std}}\,(1+\alpha^{\rm LTR}), & \hbox{for $T_1^{\rm std} > T_{\rm RH}$}.
\end{cases}
\end{equation}
Here, $\alpha^{\rm std}$ and $\alpha^{\rm LTR}$ are the values of the ratio $\rho_a^{\rm str}(T_0)/\rho_a^{\rm mis}(T_0)$ in Eq.~(\ref{proportionality}) in the standard and LTR cosmologies respectively.

\subsection{Results for LTR} \label{Results for LTR}

We now derive the regions of axion parameter space where the axion is 100$\%$ of the CDM in the LTR cosmology. We then compare them to the standard-cosmology regions.

The CDM axion parameter space in the standard cosmology depends on the PQ energy scale $f_a$ (or alternatively the axion mass $m_a$), the initial misalignment angle $\theta_i$ and the Hubble parameter during inflation $H_I$. In the LTR cosmology an additional parameter is included, the reheating temperature $T_{\rm RH}$.

If the PQ symmetry breaks after the end of inflation (Scenario I, $f_a<H_I/2\pi$), there is only one PQ scale $f_a$ for which the totality of cold dark matter is made of axions. There correspondingly is also a single value for the axion mass $m_a$. In the LTR cosmology, using the observed value for $\Omega_{\rm CDM} h^2$ in Eq.~(\ref{CDM}) and the expressions for $\Omega_a^{\rm LTR}$ in this Section, we find
\begin{equation} \label{naturalscale_LTR}
f_a^{\rm LTR} = (3.67 \pm 0.11)\times 10^{14}{\rm ~GeV}\,g_{RH}^{1/6}(T_1^{\rm LTR})\,T_{\rm RH,MeV}^{-4/3},
\end{equation}
and
\begin{equation} \label{naturalmass_LTR}
m_a^{\rm LTR} = 16.9\pm0.5\, {\rm neV}\,g_{RH}^{-1/6}(T_1^{\rm LTR})\,T_{\rm RH,MeV}^{4/3}.
\end{equation}

\begin{figure}[tb]
  \includegraphics[width=14cm]{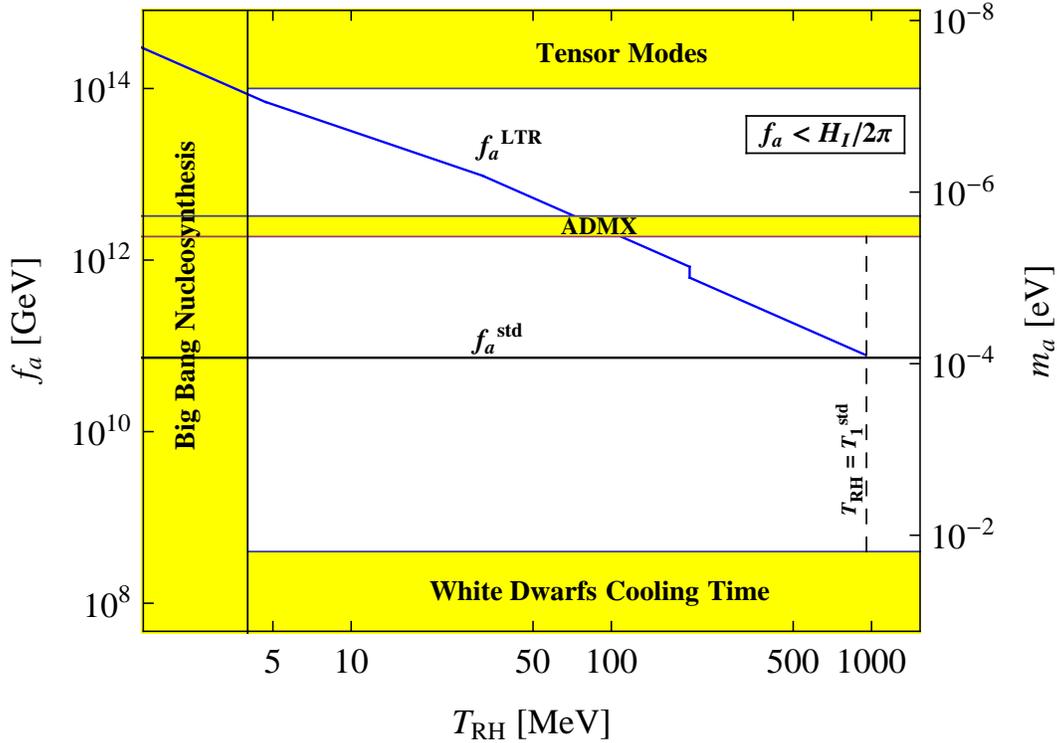}
\caption{The Peccei-Quinn scale $f_a^{\rm LTR}$ as a function of the reheating temperature $T_{\rm RH}$ for the axion to be 100$\%$ of the CDM in Scenario I ($f_a<H_I/2\pi$). Also shown are the PQ scale $f_a^{\rm std}$ in the standard cosmology, and various constraints (shaded regions).}
\label{running_Fa_LTR}
\end{figure}

In Fig.~\ref{running_Fa_LTR} we plot $f_a^{\rm LTR}$ as a function of $T_{\rm RH}$. The jumps and kinks in the $f_a^{\rm LTR}$ line are due to the different values of $g_*(T_{\rm RH})$ and $g_*(T_1^{\rm LTR})$ in Eq.~(\ref{g*}). There is also a (visually small) discontinuity between the $f_a^{\rm LTR}$ and $f_a^{\rm std}$ lines at $T_{\rm RH} = T_1^{\rm std} = T_1^{\rm LTR}$ due to different contributions from string decays. In fact, from Eqs.~(\ref{standard_density_numerical}) and~(\ref{energy_density_total_LTR}) we have
\begin{equation} \label{energy_density_total_LTR1}
\Omega^{\rm LTR}_a = \begin{cases}
1.32\,g_*^{-5/12}(T_1^{\rm std})\,\langle\theta_i^2\,f(\theta_i)\rangle\,(f_{a,12}^{\rm std})^{7/6}\,(1+\alpha^{\rm std}), & \hbox{for $T_1^{\rm std} < T_{\rm RH}$},\\
1.32\,g_*^{-5/12}(T_1^{\rm LTR})\,\langle\theta_i^2\,f(\theta_i)\rangle\,(f_{a,12}^{\rm LTR})^{7/6}\, \frac{N^{\rm LTR}}{N^{\rm std}}\,\frac{V^{\rm LTR}}{V^{\rm std}}\,(1+\alpha^{\rm LTR}), & \hbox{for $T_1^{\rm std} > T_{\rm RH}$}.
\end{cases}
\end{equation}
Equating the two lines in Eq.~(\ref{energy_density_total_LTR1}) at $T_{\rm RH} = T_1^{\rm std} = T_1^{\rm LTR}$, where $N^{\rm LTR} = N^{\rm std}$ and $V^{\rm LTR} = V^{\rm std}$, we obtain
\begin{equation} \label{condition_f_LTR}
f_a^{\rm LTR}(T_{\rm RH}\!=\!T_1^{\rm std}) = f_a^{\rm std}\,\left(\frac{1+\alpha^{\rm std}}{1+\alpha^{\rm LTR}}\right)^{6/7}.
\end{equation}
Inserting numerical values, $f_a^{\rm LTR}(T_{\rm RH}\!=\!T_1^{\rm std}) = 8.06\times 10^{10}{\rm ~GeV}$, which is slightly higher than $f_a^{\rm std}$.

In Fig.~\ref{running_Fa_LTR} we also shade out the following bounds: the bound from white dwarfs cooling times in Eq.~(\ref{whitedwarvesbound}); the indirect bound on $f_a$ from the non-detection of primordial gravitational waves arising from $f_a<H_I/2\pi$ and Eq.~(\ref{gravitywavesbound}) (region labeled ``Tensor Modes''); the bound on $T_{\rm RH}$ from Big Bang Nucleosynthesis; and the bound from the ADMX experiment \cite{asztalos, duffy} excluding a KSVZ axion with a mass $m_a$ between 1.9~$\,{\rm\mu eV}$ and 3.3~${\rm\mu eV}$. The dashed line marks the requirement that the axion starts to oscillate in the LTR cosmology, $T_{\rm RH} < T_1^{\rm std}$, with $T_1^{\rm std}$ given by Eq.~(\ref{T1natural}). The ADMX bound can be rephrased as an exclusion bound for the reheating temperature $T_{\rm RH}$. Using the expression for the axion mass in Scenario I, Eq.~(\ref{naturalmass_LTR}), the ADMX result corresponds to an exclusion of the region $72{\rm ~MeV} < T_{\rm RH} < 110{\rm ~MeV}$, valid for KSVZ axions.

Depending on $T_{\rm RH}$, $f_a^{\rm LTR}$ may differ from $f_a^{\rm std}$ in Eq.~(\ref{naturalscale}) by orders of magnitude. The maximum value of $f_a^{\rm LTR}$ is achieved for $T_{\rm RH} = 4\,{\rm MeV}$ and is, with $g_*(T_{\rm RH}) = 10.75$, $f_a^{\rm LTR} = (8.58 \pm 0.25)\times 10^{13}\,{\rm GeV}$. This value is three orders of magnitude larger than $f_a^{\rm std}$ in Eq.~(\ref{naturalscale}). As discussed in Section \ref{Discussion}, these large values of $f_a$ correspond to axion masses that are beyond the reach of current DM axion search experiments.

\begin{figure}[tb]
  \includegraphics[width=12cm]{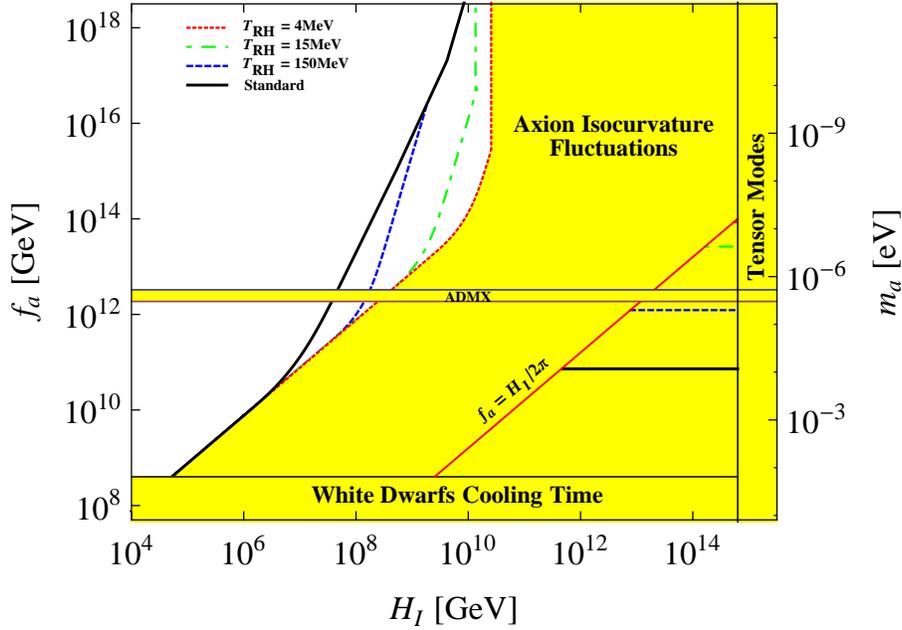}
  \caption{In the LTR cosmology, axions are the 100$\%$ of the CDM in the white region on the left (limited by a different line for each $T_{\rm RH}$) and in the narrow bands marked by horizontal lines in the lower right triangle (one line for each $T_{\rm RH}$).}
  \label{LTR}
\end{figure}

In Scenario II ($f_a>H_I/2\pi$), the parameter space is bounded by the non-detection of axion isocurvature fluctuations in the CMB spectrum, see Eq.~(\ref{isocurvature}). For $T_{\rm RH} > T_1^{\rm std} $, the isocurvature bound has the same expression as in the standard cosmology, namely Eq.~(\ref{leftmostboundary}). For $T_{\rm RH} < T_1^{\rm std}$, we eliminate $\theta_i$ in Eq.~(\ref{isocurvature}) by equating $\Omega_{\rm CDM}h^2$ with the expression for $\Omega_a^{\rm LTR}$ derived previously in this Section. The LTR isocurvature bound is then, for $T_{\rm RH} < T_1^{\rm std}$,
\begin{equation} \label{leftmostboundaryLTR}
\begin{cases}
1.05\times 10^{-2}\,g_{RH}^{1/8}(T_1^{\rm LTR})\,\sqrt{f(\theta_i)}\,T_{\rm RH,MeV}^{-1}\,f_{a,12}^{1/4}, & \hbox{for $9.2\times 10^{14}\,T_{\rm RH,MeV}^{-4/3}\,{\rm GeV} < f_a < \hat{f}_a(T_{\rm RH})$},\\
5.13 \times 10^{-2}\,T_{\rm RH,MeV}^{-1/2}, & \hbox{for $\hat{f}_a(T_{\rm RH}) < f_a$}.
\end{cases}
\end{equation}
Here $\hat{f}_a(T_{\rm RH}) = 5.69\times 10^{14}{\rm ~GeV}\,g_{RH}^{-1/2}(T_1^{\rm LTR})\,T_{\rm RH,MeV}^2$ is given by Eq.~(\ref{PQ_scale_LTR}). The LTR isocurvature bound can be approximated by
\begin{equation} \label{leftmostboundaryLTR1}
H_{I,12} = \begin{cases}
1.31\times 10^{-4}\,f_{a,12}, & \hbox{for $f_a < 9.2\times 10^{14}\,T_{\rm RH,MeV}^{-4/3}\,{\rm GeV}$},\\
1.05\times 10^{-2}\,g_{RH}^{1/8}(T_1^{\rm LTR})\,T_{\rm RH,MeV}^{-1}\,f_{a,12}^{1/4}, & \hbox{for $9.2\times 10^{14}\,T_{\rm RH,MeV}^{-4/3}\,{\rm GeV} < f_a < \hat{f}_a(T_{\rm RH})$},\\
5.13 \times 10^{-2}\,T_{\rm RH,MeV}^{-1/2}, & \hbox{for $f_a > \hat{f}_a(T_{\rm RH})$},
\end{cases}
\end{equation}
As in the case of the standard cosmology, there are two changes in the power-law dependence of $H_{I,12}$ on $f_{a,12}$ in the LTR cosmology, the first one being at $f_a = 9.2\times 10^{14}\,T_{\rm RH,MeV}^{-4/3}\,{\rm GeV}$ and the second one at $f_a = \hat{f}_a(T_{\rm RH})$. Notice that at large $f_a$, the isocurvature bound is independent of $f_a$.

When $T_{\rm RH} = T_1^{\rm std}$, the LTR and the standard isocurvature bounds coincide. This happens for
\begin{equation}
\begin{cases}
f_a = 4.6\times 10^{21}\,T_{\rm RH,MeV}^{-2}\,{\rm GeV}, & \hbox{for $T_1^{\rm std} < \Lambda$},\\
f_a = 5.3 \times 10^{28}\,T_{\rm RH,MeV}^{-6}\,{\rm GeV}, & \hbox{for $T_1^{\rm std} > \Lambda$}.
\end{cases}
\end{equation}

Fig.~\ref{LTR} shows the regions of the parameter space $(f_a,H_I)$ where the axion is 100\% of the cold dark matter in the LTR cosmology. The axion mass scale on the right is Eq.~(\ref{eq:axionmass}) with $N=1$. The region labeled ``Tensor Modes'' is excluded by the non-observation of tensor modes in the CMB fluctuations, Eq.~(\ref{gravitywavesbound}). The region labeled ``White Dwarfs Cooling Time'' is excluded from astrophysical observations of white dwarfs cooling times for KSVZ axions, Eq.~(\ref{whitedwarvesbound}) \cite{raffelt1}.
The line $f_a = H_I/2\pi$ divides the region where the PQ symmetry breaks after inflation (Scenario I, $f_a<H_I/2\pi$) from the region where it breaks during inflation (Scenario II, $f_a>H_I/2\pi$).

In the lower right region (Scenario I), the axion is the CDM particle if $f_a$ equals the value given by Eq.~(\ref{naturalscale_LTR}). Table I lists the values we plot. For comparison, we also plot the value for the standard cosmology, $f_a^{\rm std} = 7.27 \pm 0.25\times 10^{10}\,{\rm GeV}$, $m_a^{\rm std} = 85 \pm 3\, {\rm \mu eV}$ (thick line).

\begin{table}[tb]
\begin{tabular}{llll}
$T_{\rm RH}$ \hspace{4em} & $f_a^{\rm LTR}$ \hspace{12em} & $m_a^{\rm LTR}$ \hspace{6em} & Line in Fig.~1\\
\hline
$4\,{\rm MeV} $&
$(8.58 \pm 0.25)\times 10^{13} \, {\rm GeV}$ &
$ 72 \pm 2\,{\rm neV}$ &
dotted line \\
$15\,{\rm MeV} $ &
$ (2.64 \pm 0.08)\times 10^{13}{\rm ~GeV}$ &
  $235 \pm 7\,{\rm neV}$ &
dot-dashed line \\
$150\,{\rm MeV} $ &
 $(1.23 \pm 0.04)\times 10^{12}{\rm ~GeV}$ &
 $5.04 \pm 0.15 {\rm \mu eV}$ &
dashed line
\end{tabular}
\caption{Values of $f_a^{\rm LTR}$ and $m_a^{\rm LTR}$ for the axion to be 100\% of CDM in the LTR cosmology.}
\end{table}

In the upper left region (Scenario II), we plot the isocurvature bounds to the allowed parameter space for the standard cosmology (thick line) and for $T_{\rm RH} = 4{\rm ~MeV}$ (dotted line), 15 MeV (dot-dashed line) and 150 MeV (dashed line). For a given $T_{\rm RH}$, the isocurvature bound with the LTR cosmology lies below the standard line, because the entropy dilution term $\sim (T_{\rm RH}/T_1)^5$ lowers the axion energy density. Thus, more parameter space is allowed for the axion to be 100$\%$ of the CDM in the LTR cosmology than in the standard cosmology.

In the allowed region of parameter space for Scenario II, the axion can be 100$\%$ of the CDM provided the value of $\theta_i$ is chosen appropriately. This value does not depend on $H_I$, because in Scenario II $\sigma_{\theta}^2 \ll \theta_i^2$. In the standard cosmology $\theta_i$ is a function of $f_a$ only \cite{visinelli, hamann}
\begin{equation} \label{fv.theta}
f_{a,12} = \begin{cases}
\left(\frac{\Omega_{\rm CDM}h^2}{0.236\,\theta_i^2\,f(\theta_i)}\right)^{6/7}, & \hbox{for $f_a < \hat{f}_a\,\,{\rm or}\,\,\theta_i \gtrsim 10^{-3}$},\\
\left(\frac{\Omega_{\rm CDM}h^2}{0.0051\,\theta_i^2\,f(\theta_i)}\right)^{2/3}, & \hbox{for $f_a > \hat{f}_a\,\,{\rm or}\,\,\theta_i \lesssim 10^{-3}$}.
\end{cases}
\end{equation}
In the LTR cosmology, we find that $\theta_i$ depends on both $f_a$ and $T_{\rm RH}$,
\begin{equation} \label{fv.theta_LTR}
f_{a,12} = \begin{cases}
\left(\frac{\Omega_{\rm CDM}h^2}{6.35\times 10^{-7}\,\theta_i^2\,f(\theta_i)T_{\rm RH,MeV}^2}\right)^{2/3}, & \hbox{for $f_a < \hat{f}_a(T_{\rm RH})\,\,{\rm or}\,\,\theta_i \gtrsim  17\,T_{\rm RH,MeV}^{-5/2}$},\\
\left(\frac{\Omega_{\rm CDM}h^2}{7.46\times 10^{-8}\,\theta_i^2\,f(\theta_i)T_{\rm RH,MeV}}\right)^{1/2}, & \hbox{for $f_a > \hat{f}_a(T_{\rm RH})\,\,{\rm or}\,\,\theta_i \lesssim 17\,T_{\rm RH,MeV}^{-5/2}$}.
\end{cases}
\end{equation}
In Eq.~(\ref{fv.theta_LTR}) we took $g_*(T_1^{\rm LTR}) = 61.75$.

\begin{figure}[tb]
  \includegraphics[width=12cm]{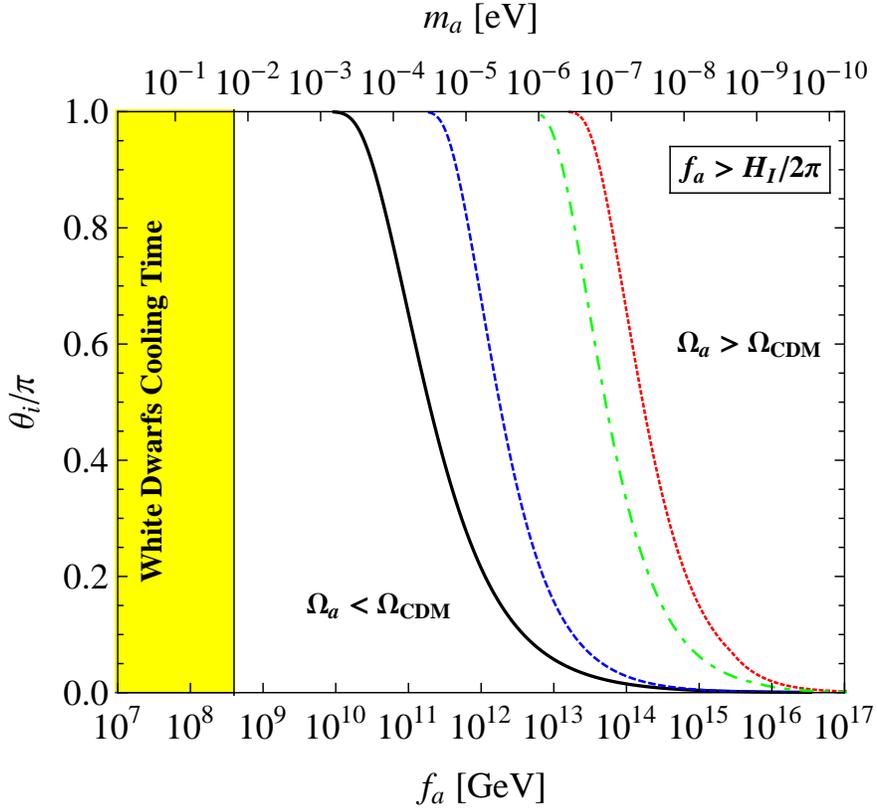}
\caption{The initial misalignment angle $\theta_i$ as a function of the Peccei-Quinn scale $f_a$ for the axion to be 100$\%$ of the CDM in Scenario II ($f_a>H_I/2\pi$): standard cosmology (black solid line), LTR cosmology with $T_{\rm RH} = 4$MeV (red dotted line), 15MeV (green dot-dashed line) or 150MeV (blue dashed line).}
\label{ThetaLTR}
\end{figure}

We plot the relation between $\theta_i$ and $f_a$ in Fig.~\ref{ThetaLTR} for the standard cosmology (thick line) and for $T_{\rm RH} = 4{\rm ~MeV}$ (dotted line), $15 {\rm ~MeV}$ (dot-dashed line) and $150 {\rm ~MeV}$ (dashed line). As $T_{\rm RH}$ decreases, one departs from the standard cosmology. The value of $\theta_i$ at fixed $f_a$, or of $f_a$ at fixed $\theta_i$, increases when $T_{\rm RH}$ decreases. The largest departure occurs at the smallest value of $T_{\rm RH}$ allowed by the BBN, $T_{\rm RH} = 4 {\rm ~MeV}$.

In all these scenarios, the value of $\theta_i$ becomes uncomfortably small at large $f_a$ if one maintains that the initial angle $\theta_i$ should not be tuned to a small value for the Peccei-Quinn mechanism to solve the strong CP problem \cite{linde}. For the sake of illustration, if we decide that $\theta_i = 0.1\pi$ is uncomfortably small, the PQ scale $f_a$ could comfortably be less than $5\times10^{12}\,{\rm GeV}$ in the standard cosmology, but could comfortably be higher in the LTR cosmologies. For example, it could comfortably be as high as $5\times 10^{15}\, {\rm GeV}$ if $T_{\rm RH} = 4{\rm ~MeV}$. This relaxes the demand of 100$\%$ axion CDM on theoretical models that prefer $f_a$ of the order of the Grand Unification (GUT) scale $\sim 10^{16}{\rm ~GeV}$.

\section{Axion CDM in the Kination cosmology} \label{Axion CDM in the Kination scenario}

We now discuss axion cold dark matter in the kination cosmology \cite{ford, kamionkowski_turner, spokoiny, joyce, tashiro, salati, profumo, rosati, pallis, chun, chung1, gomez}. The Hubble parameter for this pre-BBN cosmology is (see Ref.~\cite{chun})
\begin{equation} \label{H_kination}
H(T) = \begin{cases}
\sqrt{\frac{8\pi^3}{90}g_{\rm kin}(T)}\,\frac{T^3}{M_{Pl}T_{\rm kin}}, & {\rm for}\, T > T_{\rm kin},\\
\sqrt{\frac{8\pi^3}{90}g_*(T)}\,\frac{T^2}{M_{Pl}}, & {\rm for}\, T < T_{\rm kin}.
\end{cases}
\end{equation}
Here $T_{\rm kin}$ is the temperature at which the Universe transitions from kination domination to radiation domination, and
\begin{equation}
g_{\rm kin}(T) = \frac{g_*^2(T)}{g_*(T_{\rm kin})}.
\end{equation}
Entropy is conserved during the kination cosmology, so the scale factor during kination $a^{\rm kin}(T)$ follows the same temperature dependence as $a^{\rm std}(T)$,
\begin{equation} \label{scale_factor_kination}
g_{*S}^{1/3}(T)\,T\,a^{\rm kin}(T) = {\rm constant}.
\end{equation}
For this reason, in this Section we do not make a distinction between the scale factor in the standard and in the kination cosmologies. We write $a^{\rm kin}(T) = a^{\rm std}(T) \equiv a(T)$.

If $T_1^{\rm std} < T_{\rm kin}$, coherent oscillations in the axion field start in the radiation-dominated Universe, at the temperature $T_1^{\rm std}$ given in Eq.~(\ref{t1std}). On the contrary, if $T_1^{\rm std} > T_{\rm kin}$, coherent oscillations in the axion field start when the Universe is in its kination stage. In this case, the temperature $T_1^{\rm kin}$ at which axion oscillations begin is given by the following expression:
\begin{equation}\label{T1kin}
T_1^{\rm kin} = \begin{cases}
2.48{\rm ~GeV}\,g_{\rm kin}^{-1/6}(T_1^{\rm kin})\,T_{\rm kin,MeV}^{1/3}\,f_{a,12}^{-1/3}, & \hbox{for $T_1^{\rm kin} \lesssim \Lambda$},\\
331{\rm ~MeV}\,g_{\rm kin}^{-1/14}(T_1^{\rm kin})\,T_{\rm kin,MeV}^{1/7}\,f_{a,12}^{-1/7}, & \hbox{for $T_1^{\rm kin} \gtrsim \Lambda$}.
\end{cases}
\end{equation}
Here $T_{\rm kin,MeV} = T_{\rm kin}/{\rm MeV}$.

The axion energy density in the kination cosmology has contributions from string decays and from the misalignment mechanism.

String decays give a contribution to the present axion energy density
\begin{equation} \label{strings_kination}
\Omega_a^{\rm kin,str} = \alpha^{\rm kin}\,\Omega_a^{\rm kin,mis} = 0.42\,\bar{r}^{\rm kin}\,\Omega_a^{\rm kin,mis},
\end{equation}
where to compute $\alpha^{\rm kin}$ in Eq.~(\ref{proportionality}) we used $N_d = 1$, $\xi^{\rm kin} = 2.06$, $\zeta = 4.9$ and $\bar{r}^{\rm kin}$ is given in Eq.~(\ref{bar_r_broad}) as
\begin{equation}
\bar{r}^{\rm kin} = \frac{2}{3}\,0.8\ln(\frac{t_1}{t_{\rm PQ}}) = \frac{1.6}{3}\,\ln\left(\frac{H(f_a)}{H(T_1^{\rm kin})}\right).
\end{equation}
In the last expression we used the relation $t \propto 1/H(T)$, the fact that at the time of the PQ transition $t_{\rm PQ}$ the temperature of the Universe is $T = f_a$, and the fact that at the time $t_1$ the corresponding temperature is $T_1^{\rm kin}$. Using the expression for the kination Hubble parameter in Eq.~(\ref{H_kination}) and neglecting the term $\ln(\sqrt{\frac{g_{\rm kin}(f_a)}{g_{\rm kin}(T1^{\rm kin})}})\sim 1$, we obtain
\begin{equation}
\bar{r}^{\rm kin} = 1.6\,\ln(f_a/T_1^{\rm kin}).
\end{equation}
The temperature $T_1^{\rm kin}$ is greater than $\Lambda$ for any value of $T_{\rm RH}$ and any value of $f_a$ for which there are contributions from string decays (Scenario I, $f_a < H_I/2\pi$). Thus, using the expression for $T_1^{\rm kin}$ in the second line of Eq.~(\ref{T1kin}), we obtain
\begin{equation}
\bar{r}^{\rm kin} = 57 +\frac{16}{7}\ln f_{a,12} - \ln T_{\rm kin,MeV}.
\end{equation}
In the region of the parameters of interest for kination, $\bar{r}^{\rm kin} \sim 35$. Thus, axions from strings dominate the total axion population, the energy density $\Omega_a^{\rm kin,str}$ being one order of magnitude larger than $\Omega_a^{\rm kin,mis}$. We notice that this is opposite to what we obtained in the standard and LTR cosmologies, where the radiation of axions from axionic strings is a sub-dominant production mechanism for cold axions.

The contribution from the misalignment mechanism results from the conservation of the axion number in a comoving volume, $n_a(T) \propto a^{-3}(T)$. This gives
\begin{equation}
n_a^{\rm kin}(T_0) = \begin{cases}
n_a(T_1^{\rm std})\,\left(\frac{a(T_1^{\rm std})}{a(T_0)}\right)^3, & \hbox{for $ T_1^{\rm std} < T_{\rm kin}$},\\
n_a(T_1^{\rm kin})\left(\frac{a(T_1^{\rm kin})}{a(T_0)}\right)^3, & \hbox{for $ T_1^{\rm std} > T_{\rm kin}$}.
\end{cases}
\end{equation}
Here $n_a(T_1)$ is the function given in Eq.~(\ref{numberdensity}). One clearly has
\begin{equation}
n_a^{\rm kin}(T_0) = n_a^{\rm std}(T_0) \quad \hbox{for $T_1^{\rm std} < T_{\rm kin}$.}
\end{equation}

For $T_1^{\rm std} > T_{\rm kin}$, one obtains a different axion density. As for the LTR cosmology, we introduce the ratio between the present density $n_a^{\rm kin}(T_0)$ in the kination cosmology, and the present density $n_a^{\rm std}(T_0)$ in Eq.~(\ref{eq:nastd}) that would ensue if the cosmology were standard at temperatures $T>T_{\rm kin}$. We write, for $T_1^{\rm std} > T_{\rm kin}$,
\begin{equation}
\frac{n_a^{\rm kin}(T_0)}{n_a^{\rm std}(T_0)} =
\frac{N^{\rm kin}}{N^{\rm std}}\,\frac{V^{\rm kin}}{V^{\rm std}},
\end{equation}
where $N^{\rm kin}/N^{\rm std}$ and $V^{\rm kin}/V^{\rm std}$ are defined as follows. The ratio $N^{\rm kin}/N^{\rm std}$ is the standard-cosmology ratio of the comoving number of axions $N^{\rm kin}$  at the temperature $T_1^{\rm kin}$ to the comoving number of axions $N^{\rm std}$ at the temperature $T_1^{\rm std}$. Using Eq.~(\ref{scale_factor_kination}), we write it as
\begin{equation}
\frac{N^{\rm kin}}{N^{\rm std}} = \frac{n_a(T_1^{\rm kin})}{n_a(T_1^{\rm std})}\left(\frac{a(T_1^{\rm kin})}{a(T_1^{\rm std})}\right)^3 = \frac{n_a(T_1^{\rm kin})}{n_a(T_1^{\rm std})}\frac{g_{*S}(T_1^{\rm std})}{g_{*S}(T_1^{\rm kin})}\left(\frac{T_1^{\rm std}}{T_1^{\rm kin}}\right)^3.
\end{equation}
The ratio $V^{\rm kin}/V^{\rm std}$ is the ratio of the kination-cosmology volume $V^{\rm kin}$ to the standard-cosmology volume $V^{\rm std}$ at the temperature $T_1^{\rm kin}$,
\begin{equation}
\frac{V^{\rm kin}}{V^{\rm std}} = \left(\frac{a^{\rm kin}(T_1^{\rm kin})}{a^{\rm std}(T_1^{\rm kin})}\right)^3 = 1,
\end{equation}
The last equality follows because no significant entropy is released during the kination stage \cite{salati}, so $a^{\rm kin}(T) = a^{\rm std}(T)$.

The present axion energy density from the misalignment mechanism, in units of the critical density, is therefore
\begin{equation} \label{compare_kin}
\Omega_a^{\rm kin,mis} = \begin{cases}
\Omega_a^{\rm std,mis}, & \hbox{for $T_1^{\rm std} < T_{\rm kin}$},\\
\Omega_a^{\rm std,mis}\,\frac{N^{\rm kin}}{N^{\rm std}}, & \hbox{for $T_1^{\rm std} > T_{\rm kin}$}.
\end{cases}
\end{equation}

Inserting numerical values, the first line of Eq.~(\ref{compare_kin}) is given by Eq.~(\ref{standard_density_numerical}), while the second line reads
\begin{equation}\label{density_kination}
\Omega_a^{\rm kin,mis} h^2 = 1150 \,g_*^{-1/2}(T_{\rm kin})\,\langle\theta_i^2\,f(\theta_i)\rangle\,f_{a,12}\,T_{\rm kin,MeV}^{-1}.
\end{equation}
Due to the peculiar dependence of the Hubble rate with temperature in kination, $H(T) \sim T^3$, there is no distinction in Eq.~(\ref{density_kination}) between $\Omega_a^{\rm kin,mis}$ for $T_1^{\rm kin} \gtrsim \Lambda$ and for $T_1^{\rm kin} \lesssim \Lambda$.

Finally, the present axion energy density in the kination cosmology is given by the sum of the misalignment mechanism and the string decay contributions
\begin{equation} \label{energy_density_total_kination}
\Omega_a^{\rm kin} = \Omega_a^{\rm kin,mis} + \Omega_a^{\rm kin,str} = \begin{cases}
\Omega_a^{\rm std,mis}\,(1+\alpha^{\rm std}), & \hbox{for $T_1^{\rm std} < T_{\rm kin}$},\\
\Omega_a^{\rm std,mis} \frac{N^{\rm kin}}{N^{\rm std}}\,(1+\alpha^{\rm kin}), & \hbox{for $T_1^{\rm std} > T_{\rm kin}$}.
\end{cases}
\end{equation}
Here, $\alpha^{\rm std}$ and $\alpha^{\rm kin}$ are the values of the ratio $\rho_a^{\rm str}(T_0)/\rho_a^{\rm mis}(T_0)$ in Eq.~(\ref{proportionality}) in the standard and kination cosmologies respectively.

\subsection{Results for kination} \label{Results for kination}

We now derive the regions of the axion parameter space where the axion is 100$\%$ of the CDM in the kination cosmology. We then compare them to the standard-cosmology regions.

The axion parameter space in kination cosmology depends on $f_a$, $H_I$, $\theta_i$, and the additional parameter $T_{\rm kin}$.

If the PQ symmetry breaks after the end of inflation (Scenario I, $f_a<H_I/2\pi$), there is only one PQ scale $f_a$ for which the totality of cold dark matter is made of axions. There correspondingly is also a single value of the axion mass $m_a$. In the kination cosmology, using the observed value of $\Omega_{\rm CDM} h^2$ in Eq.~(\ref{CDM}), and the expressions for $\Omega_a^{\rm kin}$ derived in this Section, we find
\begin{equation} \label{naturalscale_kination}
f_a^{\rm kin} = (7.9 \pm 0.2) \times 10^6\,{\rm GeV}\,g_*^{1/2}(T_{\rm kin})\,\frac{T_{\rm kin,MeV}}{57 +\frac{16}{7}\ln f_{a,12}^{\rm kin} - \ln T_{\rm kin,MeV}},
\end{equation}
and
\begin{equation}
m_a^{\rm kin} = 739\pm22\, {\rm meV}\,g_*^{-1/2}(T_{\rm kin})\,\bar{r}^{\rm kin}\,T^{-1}_{\rm kin,MeV}.
\end{equation}
In Eq.~(\ref{naturalscale_kination}) we used the explicit expression for $\bar{r}^{\rm kin}$ derived in Section~\ref{Axions from string decays}.

\begin{figure}[tb]
  \includegraphics[width=14cm]{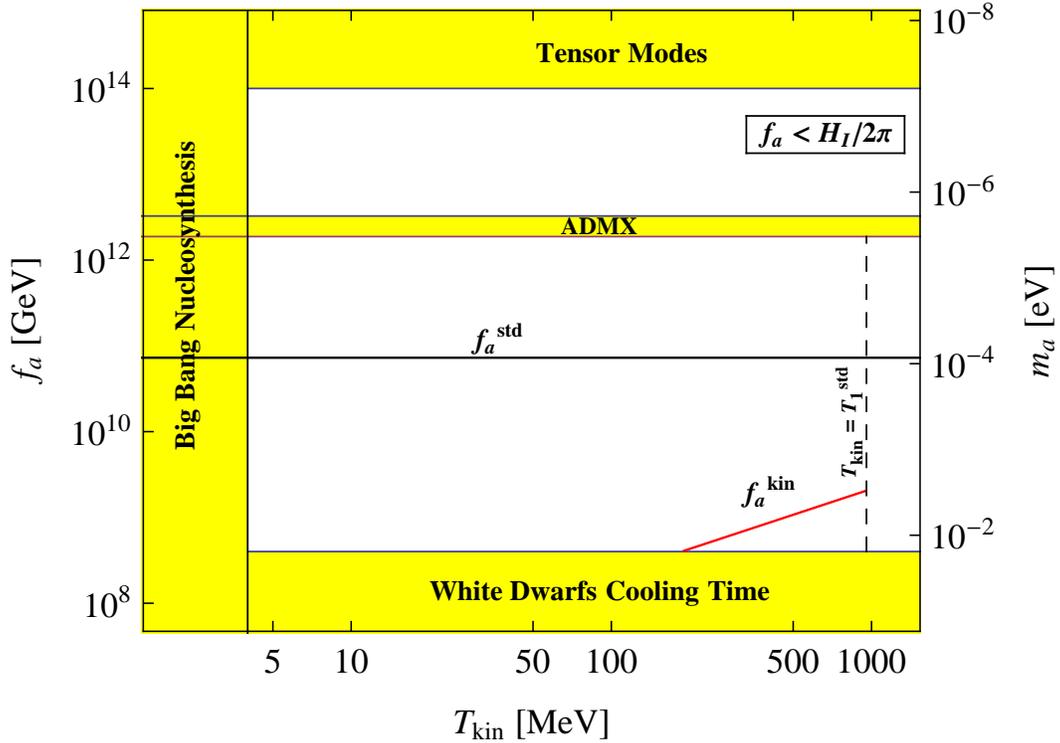}
\caption{The Peccei-Quinn scale $f_a^{\rm kin}$ as a function of the kination temperature $T_{\rm kin}$ for the axion to be 100$\%$ of the CDM in Scenario I ($f_a<H_I/2\pi$). Also shown are the PQ scale $f_a^{\rm std}$ in the standard cosmology, and various constraints (shaded regions).}
\label{running_Fa_kination}
\end{figure}

In Fig.~\ref{running_Fa_kination} we plot $f_a^{\rm kin}$ as a function of $T_{\rm kin}$. The function $f_a^{\rm kin}$ does not present jumps, because both $g_*(T_1^{\rm kin})$ and $g_*(T_{\rm kin})$ do not change in the domain of $f_a^{\rm kin}$. The discontinuity between the $f_a^{\rm kin}$ and $f_a^{\rm std}$ lines at $T_{\rm kin} = T_1^{\rm std} = T_1^{\rm kin}$ is due to different contributions from string decays. In fact, from Eqs.~(\ref{standard_density_numerical}) and~(\ref{energy_density_total_kination}) we have
\begin{equation} \label{energy_density_total_kination1}
\Omega^{\rm kin}_a = \begin{cases}
1.32\,g_*^{-5/12}(T_1^{\rm std})\,\langle\theta_i^2\,f(\theta_i)\rangle\,(f_{a,12}^{\rm std})^{7/6}\,(1+\alpha^{\rm std}), & \hbox{for $T_1^{\rm std} < T_{\rm kin}$},\\
1.32\,g_*^{-5/12}(T_1^{\rm kin})\,\langle\theta_i^2\,f(\theta_i)\rangle\,(f_{a,12}^{\rm kin})^{7/6}\, \frac{N^{\rm kin}}{N^{\rm std}}\,\frac{V^{\rm kin}}{V^{\rm std}}\,(1+\alpha^{\rm kin}), & \hbox{for $T_1^{\rm std} > T_{\rm kin}$}.
\end{cases}
\end{equation}
Equating the two lines in Eq.~(\ref{energy_density_total_kination1}) at $T_{\rm kin} = T_1^{\rm std} = T_1^{\rm kin}$, where $N^{\rm kin} = N^{\rm std}$, we obtain
\begin{equation} \label{condition_f_kination}
f_a^{\rm kination}(T_{\rm kin}\!=\!T_1^{\rm std}) = f_a^{\rm std}\,\left(\frac{1+\alpha^{\rm std}}{1+\alpha^{\rm kin}}\right)^{6/7}.
\end{equation}
We find $f_a^{\rm kin}(T_{\rm kin}\!=\!T_1^{\rm std}) = 2.04\times 10^9{\rm ~GeV}$.

In Fig.~\ref{running_Fa_kination} we also shade out the following bounds: the bound from white dwarfs cooling times in Eq.~(\ref{whitedwarvesbound}); the indirect bound on $f_a$ from the non-detection of primordial gravitational waves arising from $f_a<H_I/2\pi$ and Eq.~(\ref{gravitywavesbound}) (region labeled ``Tensor Modes''); the bound on $T_{\rm kin}$ from Big Bang Nucleosynthesis; and the bound from the ADMX experiment excluding a KSVZ axion with a mass $m_a$ between 1.9~$\,{\rm\mu eV}$ and 3.3~${\rm\mu eV}$. The dashed line marks the requirement that the axion starts to oscillate in the kination cosmology, $T_{\rm kin} < T_1^{\rm std}$, with $T_1^{\rm std}$ given by Eq.~(\ref{T1natural}).

The PQ scale $f_a^{\rm kin}$ is orders of magnitude lower than the PQ scale $f_a^{\rm std}$ in the standard cosmology. The low values of $f_a^{\rm kin}$ in comparison with $f_a^{\rm std}$ is due to two different reasons. The first reason is that, since coherent oscillations of the axion field start later in the kination cosmology than in the standard cosmology, the initial comoving number of axions $N^{\rm kin}$ is higher than $N^{\rm std}$. The second reason is that the contribution from axionic strings to $\Omega_a^{\rm kin}$ in the kination cosmology is much higher than the same contribution to $\Omega_a^{\rm std}$ in the standard cosmology. Then, at a given PQ scale $f_a$, the energy density $\Omega_a^{\rm kin} > \Omega_a^{\rm std}$. A lower PQ scale is thus required in order to have the same CDM energy density $\Omega_{\rm CDM}$.

The PQ scale $f_a^{\rm kin}$ can be so small as to violate the limit from the white dwarfs cooling time in Eq.~(\ref{whitedwarvesbound}). This imposes the requirement $T_{\rm kin} > 217 {\rm ~MeV}$ if axions are 100$\%$ of the CDM. This requirement is more stringent than the BBN constraint $T_{\rm kin} > 4 {\rm ~MeV}$.

In Scenario II ($f_a>H_I/2\pi$), the parameter space is bounded by the non-detection of axion isocurvature fluctuations in the CMB spectrum, Eq.~(\ref{isocurvature}). For $T_{\rm kin} > T_1^{\rm std}$, the isocurvature bound has the same expression, Eq.~(\ref{leftmostboundary}), as in the standard cosmology. For $T_{\rm kin} < T_1^{\rm std}$, we eliminate $\theta_i$ in Eq.~(\ref{isocurvature}) by equating $\Omega_{\rm CDM}$ with the expression for $\Omega_a^{\rm kin}$ derived in this Section. The resulting kination isocurvature bound for $T_{\rm kin} < T_1^{\rm std}$ is
\begin{equation} \label{leftmostboundarykination}
H_{I,12} < 7.48\times 10^{-7}\,\sqrt{f(\theta_i)}\,f_{a,12}^{1/2}\,T_{\rm kin,MeV}^{1/2}.
\end{equation}
This bound can be approximated by
\begin{equation} \label{leftmostboundarykination1}
H_{I,12} = \begin{cases}
1.31\times 10^{-4}\,f_{a,12}, & \hbox {for $f_a < 3.26\times 10^7{\rm ~GeV}\,T_{\rm kin,MeV}$},\\
7.48\times 10^{-7}\,f_{a,12}^{1/2}\,T_{\rm kin,MeV}^{1/2}, & \hbox{for $f_a > 3.26\times 10^7{\rm ~GeV}\,T_{\rm kin,MeV}$}.
\end{cases}
\end{equation}
Contrary to the cases of standard and LTR cosmologies, in the kination cosmology there is only one change in the power-law dependence of $H_{I,12}$ on $f_{a,12}$, namely at $f_a =  3.26\times 10^7{\rm ~GeV}\,T_{\rm kin,MeV}$. This change is due to the effects of anharmonicities.

When $T_{\rm kin} = T_1^{\rm std}$, the kination and the standard isocurvature bounds coincide. This happens for
\begin{equation}
\begin{cases}
f_a = 4.6\times 10^{21}\,T_{\rm kin,MeV}^{-2}\,{\rm GeV}, & \hbox{for $T_1^{\rm std} < \Lambda$},\\
f_a = 5.3 \times 10^{28}\,T_{\rm kin,MeV}^{-6}\,{\rm GeV}, & \hbox{for $T_1^{\rm std} > \Lambda$}.
\end{cases}
\end{equation}

\begin{figure}[tb]
  \includegraphics[width=13cm]{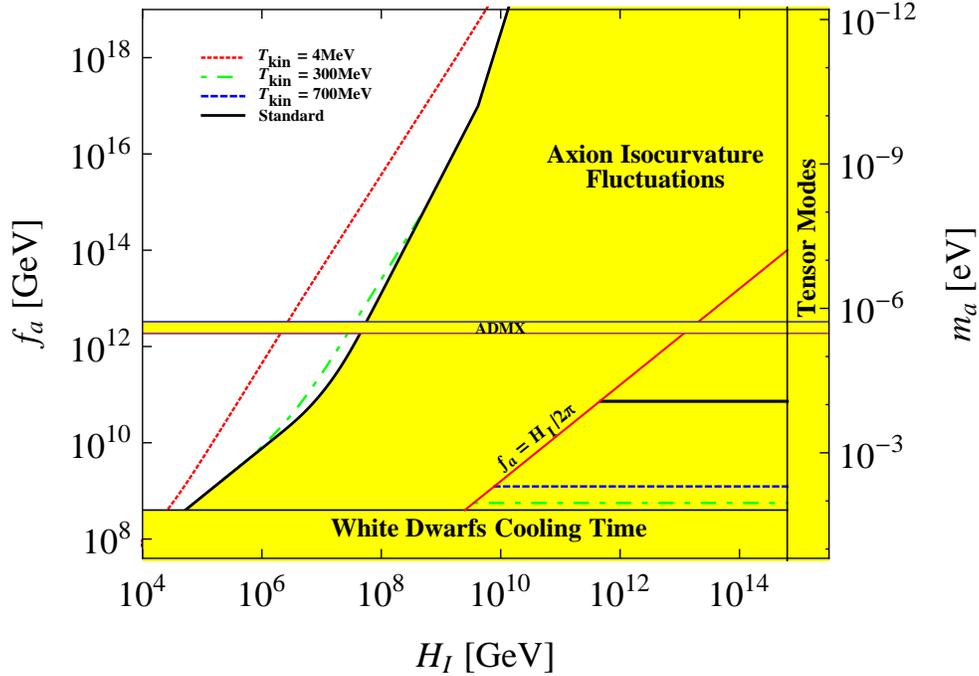}
\caption{In the kination cosmology, axions are the 100$\%$ of the CDM in the white region on the left (limited by a different line for each $T_{\rm kin}$) and in the narrow bands marked by horizontal lines in the lower right triangle (one line for each $T_{\rm kin}$).}
\label{kination}
\end{figure}

Fig.~\ref{kination} shows the regions of the parameter space $(f_a,H_I)$ where the axion is 100$\%$ of the cold dark matter in the kination cosmology. The axion mass scale on the right is Eq.~(\ref{eq:axionmass}) with $N=1$. The region labeled ``Tensor Modes'' is excluded by the non-observation of tensor modes in the CMB fluctuations, Eq.~(\ref{gravitywavesbound}). The region labeled ``White Dwarfs Cooling Time'' is excluded from astrophysical observations of white dwarfs cooling times for KSVZ axions, Eq.~(\ref{whitedwarvesbound}). A similar bound from supernovae applies to other axion models \cite{raffelt}. The line $f_a = H_I/2\pi$ divides the region where the PQ symmetry breaks after inflation (Scenario I, $f_a<H_I/2\pi$) from the region where it breaks during inflation (Scenario II, $f_a>H_I/2\pi$).

In the lower right region (Scenario I), the axion is the CDM particle if $f_a$ equals the value given by Eq.~(\ref{naturalscale_kination}). Table II lists the values we plot. Notice that the line at $T_{\rm kin} = 4 {\rm ~MeV}$ does not appear in this figure because it is excluded by the bound from white dwarfs cooling times. For comparison, we also plot the value for the standard cosmology, $f_a^{\rm std} = 7.27 \pm 0.25\times 10^{10}\,{\rm GeV}$, $m_a^{\rm std} = 85 \pm 3\, {\rm \mu eV}$ (thick line).

\begin{table}[tb]
\begin{tabular}{llll}
$T_{\rm kin}$ \hspace{4em} & $f_a^{\rm kin}$ \hspace{12em} & $m_a^{\rm kin}$ \hspace{6em} & Line in Fig.~\ref{kination}\\
\hline
$4\,{\rm MeV} $ &
$(3.8\pm 0.1)\times 10^6 \, {\rm GeV}$ &
$ 1.63 \pm 0.05\,{\rm eV}$ &
dotted line \\
$300\,{\rm MeV} $ &
$(5.45 \pm 0.2)\times 10^8{\rm ~GeV}$ &
  $11.4 \pm 0.4{\rm meV}$ &
dot-dashed line \\
$700\,{\rm MeV} $ &
 $(1.24 \pm 0.04)\times 10^9{\rm ~GeV}$ &
 $5.0 \pm 0.2 {\rm meV}$ &
dashed line
\end{tabular}
\caption{Values of $f_a^{\rm kin}$ and $m_a^{\rm kin}$ for the axion to be 100\% of CDM in the kination cosmology.}
\end{table}

\begin{figure}[tb]
  \includegraphics[width=12cm]{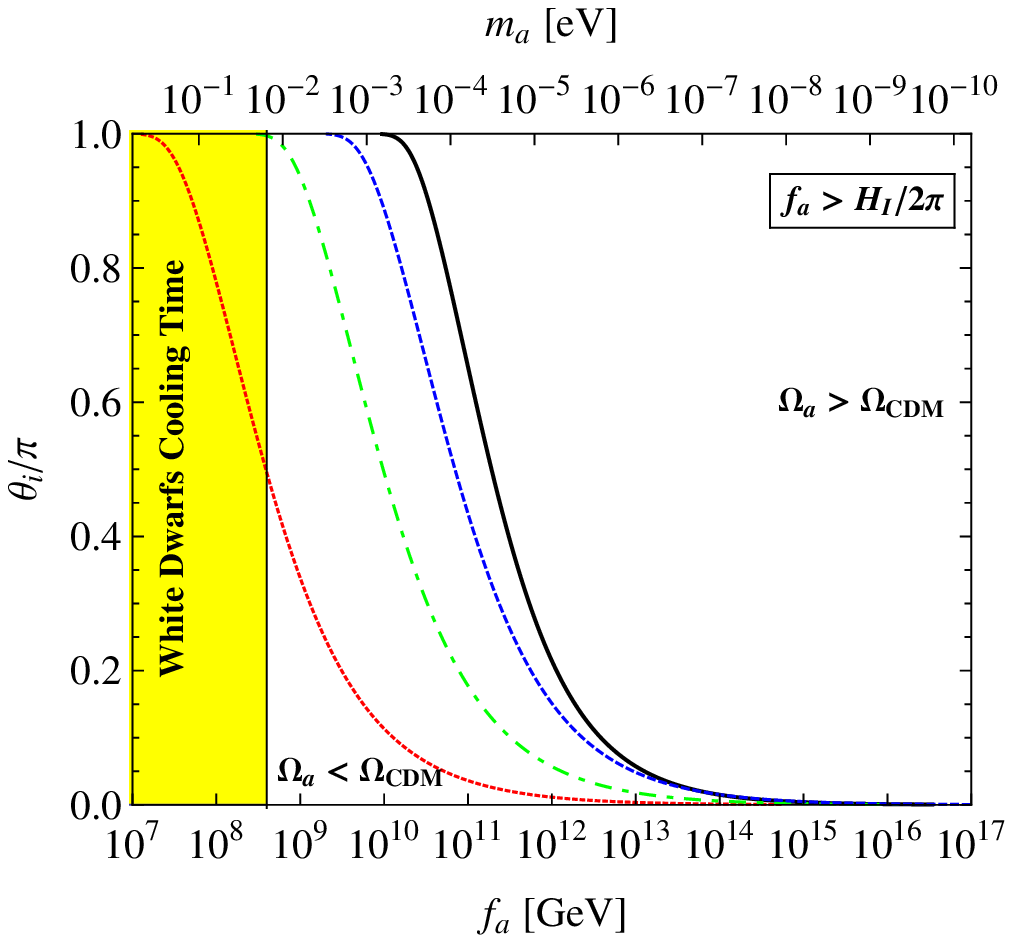}
\caption{The initial misalignment angle $\theta_i$ as a function of the Peccei-Quinn scale $f_a$ for the axion to be 100$\%$ of the CDM in Scenario II ($f_a>H_I/2\pi$): standard cosmology (black solid line), kination cosmology with $T_{\rm kin} = 4$MeV (red dotted line), 300MeV (green dot-dashed line) or 700MeV (blue dashed line).}
\label{Thetakination}
\end{figure}

In the upper left region (Scenario II), we plot the isocurvature bounds to the allowed parameter space for the standard cosmology (thick line) and for $T_{\rm kin} = 4{\rm ~MeV}$ (dotted line), 300 MeV (dot-dashed line) and 700 MeV (dashed line). For a given $T_{\rm kin}$, the kination isocurvature bound lies above the standard line. This is due to the fact that $N^{\rm kin} > N^{\rm std}$. Thus, less parameter space is allowed for the axion to be 100$\%$ of the CDM in the kination cosmology than in the standard cosmology.

In the allowed region of parameter space for Scenario II, the axion can be 100$\%$ of the CDM provided the value of $\theta_i$ is chosen appropriately. This value does not depend on $H_I$, because in Scenario II $\sigma_{\theta}^2 \ll \theta_i^2$. In the kination cosmology we find that $\theta_i$ depends on both $f_a$ and $T_{\rm kin}$ as
\begin{equation} \label{fv.theta_kination}
f_{a,12} = \frac{\Omega_{\rm CDM}h^2\,g_*^{1/2}(T_{\rm kin})\,T_{\rm kin,MeV}}{1150\,\theta_i^2\,f(\theta_i)}.
\end{equation}
We plot this relation between $f_a$ and $\theta_i$ in Fig.~\ref{Thetakination} for the standard cosmology (thick line) and for $T_{\rm kin} = 4{\rm ~MeV}$ (dotted line), $300 {\rm ~MeV}$ (dot-dashed line) and $700 {\rm ~MeV}$ (dashed line). As $T_{\rm kin}$ decreases, one departs from the standard cosmology. The value of $\theta_i$ at fixed $f_a$, or of $f_a$ at fixed $\theta_i$, decreases when $T_{\rm kin}$ decreases. This is opposite to the behavior in the LTR cosmology. The largest departure occurs at the smallest value of $T_{\rm kin}$ allowed by the BBN, $T_{\rm kin} = 4 {\rm ~MeV}$.

The problem of $\theta_i$ values uncomfortably small to solve the strong CP problem is more severe in the kination cosmology than in the standard cosmology.

\section{Discussion} \label{Discussion}

\subsection{Comparison to previous work}

Axions in kination cosmology were studied only in Ref.~\cite{kamionkowski} and only as hot dark matter (i.e.\ thermally produced in the hot primordial soup). To the extent of our knowledge, cold dark matter axions in kination cosmology were not examined before.

Axions in the LTR cosmology were studied before \cite{dine, steinhardt, lazarides, yamamoto, kawasaki, riotto}, but only to determine the cosmological bound on the PQ scale in what we call Scenario I, namely $f_a<H_I/2\pi$, in which the Peccei-Quinn symmetry breaks after the end of inflation. Our work can also be used to set an upper bound on the PQ scale by imposing
\begin{equation}
\Omega_a\,h^2 < \Omega_{CDM}\,h^2.
\end{equation}
These bounds can be read off the figures and the equations in this paper, all of which represent the equation $\Omega_a\,h^2 < \Omega_{CDM}\,h^2$. We remark that therefore our work extends previous papers in that we have examined also the region $f_a>H_I/2\pi$, where the PQ symmetry breaks during inflation, have updated $\Omega_{\rm CDM}$ to the current observational value, have used an improved constraint on $T_{\rm RH}$ from Big Bang Nucleosynthesis, and have included anharmonicities in the axion potential.

Our numerical result for the highest allowed value of the PQ scale in the LTR cosmology $f_a^{\rm LTR} = 8.58\times 10^{13} {\rm ~GeV}$, obtained for $T_{\rm RH} = 4{\rm ~MeV}$, differs from previous authors. Steinhardt and Turner \cite{steinhardt} showed that the entropy production due to the decay of a massive scalar field raises the maximum PQ scale to $f_a^{\rm LTR} \sim 10^{18} {\rm ~GeV}$, but they were corrected by Kawasaki {\it et al.} \cite{kawasaki} for using the value of $T_1$ in the standard cosmology instead of the LTR cosmology. Kawasaki {\it et al.} \cite{kawasaki} used $T_1$ in the LTR cosmology and obtained $f_a^{\rm LTR} \sim 10^{15} {\rm ~GeV}$; however, they used $\Omega_a\,h^2 = 1$ and $T_{\rm RH} = 1{\rm ~MeV}$ instead of the current value $\Omega_{\rm CDM}h^2 = 0.1131\pm 0.0034$ and the current BBN bound $T_{\rm RH}>4{\rm ~MeV}$. Giudice {\it et al.} \cite{riotto} find for the maximum PQ scale in the LTR cosmology the value $f_a^{\rm LTR} \sim 10^{16} {\rm ~GeV}$, which is higher than ours for the same reasons as for Kawasaki {\it et al.}

\subsection{Effects of changing the string decay parameters}

The computation of $f_a^{\rm std}$ in Eq.~(\ref{naturalscale}), $f_a^{\rm LTR}$ in Eq.~(\ref{naturalscale_LTR}) and $f_a^{\rm kin}$ in Eq.~(\ref{naturalscale_kination}) strongly relies on the model used to describe the axionic string evolution and the energy spectrum of emitted axions.

In Section \ref{Axions from string decays} we discussed the dependence of $\bar{r}$ and $\xi$ on the model for the axionic string oscillation and radiation spectrum. There we showed how these quantities in a modified cosmological scenario are related to their values in the standard cosmology. For the latter, we used the values $\bar{r}^{\rm std} = 0.8$ and $\xi^{\rm std} = 1$, obtained assuming that an axionic string radiates axions in a broad energy spectrum \cite{wall_solution, harari, hagmann, hagmann1} and that the axionic string network is a global string network \cite{wall_solution, hagmann, hagmann, string_simulation}. In the following we refer to these assumptions as Model A. With these values, the axionic string contribution to the total axion energy density in the standard and LTR cosmologies is subdominant compared to the contribution from the misalignment mechanism, while it is dominant in the kination cosmology.

We now discuss the modification to the axion parameter space when we assume that axionic strings radiate axions in a narrow energy spectrum \cite{davis, davis2, battye, shellard}, and that the axionic strings network is a local strings network \cite{local_strings_results}. In this case, $\bar{r}^{\rm std} = 70$ and $\xi^{\rm std} = 13$. We call this set of assumptions Model B. With these values, the contribution from strings to the axion energy density in the standard cosmology is dominant ($N_d=1$, $\zeta = 4.9$),
\begin{equation}
\Omega_a^{\rm std,str} \sim 200\,\Omega_a^{\rm std,mis}.
\end{equation}
This affects the value of $\Omega_a^{\rm std} = \Omega_a^{\rm std,str} + \Omega_a^{\rm std,mis}$, which in Model B is about two hundred times higher than in Model A. As a consequence, with Model B, the value of the PQ scale $f_a^{\rm std}$ for which $\Omega_a^{\rm std} = \Omega_{\rm CDM}$ in Scenario I is
\begin{equation}
f_a^{\rm std} = 9.3 \times 10^8 {\rm ~GeV}.
\end{equation}
This value is smaller than that in Eq.~(\ref{naturalscale}), obtained with Model A. It is of the order of the astrophysical constraint from white dwarfs cooling times in Eq.~(\ref{whitedwarvesbound}).

We conclude that, depending on the model for the axionic string and its emission spectrum, $f_a^{\rm std}$ in the standard cosmology can range from the value $(7.27 \pm 0.25) \times 10^{10}\,{\rm GeV}$ in Eq.~(\ref{naturalscale}) to the astrophysical bound from white dwarfs cooling times $4\times 10^{8}{\rm ~GeV}$ (the errors come from the error on $\Omega_{CDM}$ only). Correspondingly, $m_a^{\rm std}$ can range from $85\pm3{\rm ~\mu eV}$ to $15 \pm 1{\rm ~meV}$.

In non-standard cosmologies, when going to Model B, we must redo the calculation of $\bar{r}$ and $\xi$ using the formulas in Section \ref{Axions from string decays}. For the Model B value $\xi^{\rm std} = 13$, Eq.~(\ref{c}) gives $c =(\sqrt{52}+1)/52 \simeq 0.158$ and Eq.~(\ref{gamma_0}) gives
\begin{equation}
\xi \sim 10\left(2-3\beta+2\sqrt{2.41\beta^2-3\beta+1}\right)^2\quad\hbox{(for $\xi^{\rm std} = 13$)}.
\end{equation}
The values of $\xi$ in Model B are then $\xi^{\rm LTR} = 2.8$ for the LTR cosmology ($\beta = 2/3$)and $\xi^{\rm kin} = 41$ for the kination cosmology ($\beta = 1/3$).

For the parameter $\bar{r}$, we turn to Eq.~(\ref{bar_r1}). The results from Model A favor the fast-oscillating axionic strings model, which predicts $\bar{r}$ as given in Eq.~(\ref{bar_r_broad}). Model B points toward a slow-oscillating axionic string for which $\bar{r}$ is given by Eq.~(\ref{bar_r_sharp}). Using Model B, for the illustrative case $\delta = (10^{12}{\rm ~GeV})^{-1}$ and $T_{\rm RH} = T_{\rm kin} = 4\,{\rm MeV}$, we obtain $\bar{r} \simeq 70$ in the standard cosmology, $\bar{r}^{\rm LTR} \simeq 20$ in the LTR cosmology and $\bar{r}^{\rm kin} \simeq 4300$ in the kination cosmology.

The LTR and kination axion energy densities from axionic strings in Model B are then, with $N_d = 1$ and $\zeta = 4.9$,
\begin{equation}
\Omega_a^{\rm LTR,str}h^2 = 11.5\,\Omega_a^{\rm LTR,mis}\quad\hbox{and}\quad\Omega_a^{\rm kin,str}h^2 = 3.6\times10^4\,\Omega_a^{\rm kin,mis}.
\end{equation}
The higher axionic string contributions in Model B with respect to Model A sensibly lower the values of the PQ scales $f_a^{\rm LTR}$ and $f_a^{\rm kin}$ for which the axion is 100$\%$ of the CDM. We have, taking $T_{\rm RH} = T_{\rm kin} = 4\,{\rm MeV}$,
\begin{equation}
f_a^{\rm LTR} = (1.43\pm 0.04)\times 10^{13}{\rm ~GeV}\quad\hbox{and}\quad f_a^{\rm kin} = (4.3 \pm 0.1)\times 10^3{\rm ~GeV}.
\end{equation}

\subsection{Distinguishing non-standard cosmologies observationally}

Here we discuss how one might be able to distinguish different non-standard cosmologies before BBN using properties of the axion cold dark matter population.

One may try to distinguish non-standard cosmologies by measuring both the axion CDM density $\Omega_{\rm CDM}$ and the axion mass $m_a$. However, one immediately runs into the following problem.

Assume, for example, that the axion is found to be the main CDM component and the axion mass is measured at $m_a \simeq 10^{-3}{\rm ~eV}$. These facts can be ascribed to two different cosmological models. The first model involves the axion field evolving in the standard cosmology, with the dominant contribution to the total axion energy density coming from axionic strings and only a tiny fraction from the misalignment mechanism, as Model B would predict. The second model involves a stage of kination before BBN lasting until $T_{\rm kin} \sim 900 {\rm ~MeV}$, with the contribution from axionic strings and from the misalignment mechanism of the same order of magnitude, as in Model A.

These uncertainties in the production of axion from strings decay prevent distinguishing non-standard cosmologies with this method alone.

One may complement the measurements of $\Omega_{\rm CDM}$ and $m_a$ with a measurement of the axion CDM velocity dispersion $\delta v$. The latter allows non-standard cosmologies to be distinguished, at least in principle. The argument proceeds as follows.

When axions start to oscillate at the temperature $T_1$, axions from vacuum realignment and axionic string decay have momentum dispersion of order the Hubble scale at $T_1$ \cite{sikivie},
\begin{equation} \label{momentum_dispersion}
\delta p(T_1) \simeq H(T_1).
\end{equation}
The momentum dispersion scales with the scale factor as $\delta p(T) \propto 1/a(T)$. In the standard cosmology, the velocity dispersion at present is then
\begin{equation}
\delta v^{\rm std} \simeq \frac{H(T_1^{\rm std})}{m_a}\left(\frac{a(T_1^{\rm std})}{a(T_0)}\right) = \left(\frac{\rm\mu eV}{m_a}\right)^{5/6}\,1.4\times10^{-8}{\rm ~m/s}.
\end{equation}

In the kination cosmology,
\begin{equation}
\delta v^{\rm kin} \simeq \frac{H(T_1^{\rm kin})}{m_a}\left(\frac{a(T_1^{\rm kin})}{a(T_0)}\right) = T_{\rm kin,MeV}^{-5/7}\,\left(\frac{\rm\mu eV}{m_a}\right)^{5/7}\,8.5\times 10^{-7}{\rm ~m/s}.
\end{equation}
It is clear that if one has measured $m_a$, a measurement of $\delta v$ will give the value of $T_{\rm kin}$.

Similarly, in the LTR cosmology,
\begin{equation}
\delta v^{\rm LTR} \simeq \frac{H(T_1^{\rm LTR})}{m_a}\left(\frac{a(T_1^{\rm LTR})}{a(T_0)}\right) = \left(\frac{\rm\mu eV}{m_a}\right)^{5/6}\,7.6\times 10^{-9}{\rm ~m/s}.
\end{equation}

A difficulty in measuring $\delta v$ may arise from virialization of the axion population within galactic dark halos, although it has been claimed that $\delta v$ would be preserved in the phase-space evolution \cite{sikivie}.

\section{Conclusions} \label{Conclusions}

In this paper we have examined the parameter regions in which the axion is 100\% of the cold dark matter density in cosmologies that are non-standard before Big Bang nucleosynthesis. We have recognized two ways in which these regions change in going from the standard cosmology to the non-standard cases. If the Peccei-Quinn symmetry breaks after the end of inflation (Scenario I), the axion CDM regions shift to different values of the axion mass $m_a$ (or of the corresponding PQ scale $f_a$). If the PQ symmetry breaks during inflation (Scenario II), the axion CDM regions can shrink or expand according to the cosmological model.

We have considered two different non-standard cosmologies that change the axion CDM regions in opposite directions. In the low temperature reheating (LTR) cosmology, the axion CDM regions shift to lower axion masses in Scenario I and expand in Scenario II. In the kination cosmology, the axion CDM regions shift to higher axion masses in Scenario I and shrink in Scenario II.

Different axionic string models lead to different quantitative results, but the overall modifications from the standard cosmology follow the same trend.

We have also commented on the possibility to distinguish standard and non-standard cosmologies using observable properties of the axion CDM population. We have tentatively concluded that the axion velocity dispersion may be a good indicator of the cosmology before Big Bang nucleosynthesis.

\begin{acknowledgments}
This work was partially supported by NSF Grant No. PHY-0456825.
\end{acknowledgments}


\begin{thebibliography}{100}

\bibitem{hannestad1} M. Kawasaki, K. Kohri, and N. Sugiyama, Phys.\ Rev.\ Lett.\ {\bf 82}, 4168 (1999); Phys.\ Rev.\ D {\bf 62}, 023506 (2000); S.~Hannestad, Phys.\ Rev.\ D {\bf 70}, 043506 (2004).
\bibitem{decay_inflaton} A.~D.~Dolgov, A.~D.~Linde, Phys.\ Lett.\ B {\bf 116}, 329 (1982).
\bibitem{decay_inflaton1} L.~F.~Abbott, E.~Fahri, M.~Wise, Phys.\ Lett.\ B {\bf 117}, 29 (1982).
\bibitem{parametric_resonance} J.~H.~Traschen, R.~H.~Brandenberger, Phys.\ Rev.\ D {\bf 42}, 2491 (1990); A.~D.~Dolgov, D.~P.~Kirilova Sov.\ J.\ Nucl.\ Phys.\ {\bf 51}, 172 (1990); L.~Kofman, A.~D.~Linde, A.~A.~Starobinsky, Phys.\ Rev.\ Lett.\ {\bf 73}, 3195 (1994).
\bibitem{low_reheat_inflation} M.~Bastero-Gil, S.~F.~King, Phys.\ Rev.\ D {\bf 63}, 123509 (2001); A.~Kudo, M.~Yamaguchi, Phys.\ Lett.\ B {\bf 516}, 151 (2001).
\bibitem{turner_LTR} M.~S.~Turner, Phys.\ Rev.\ D {\bf 28}, 1243 (1983); R.~J.~Sherrer, M.~S.~Turner, {\it ibid.} {\bf 31}, 681 (1985).
\bibitem{dine} M.~Dine, W.~Fishler, Phys.\ Lett.\ B {\bf 120}, 137 (1983).
\bibitem{steinhardt} P.~J.~Steinhardt, M.~S.~Turner, Phys.\ Lett.\ B {\bf 129}, 51 (1983).
\bibitem{peccei} R.~D.~Peccei, H.~R.~Quinn, Phys. Rev. Lett. {\bf 38}, 1440 (1977); Phys.\ Rev.\ D {\bf 16}, 1791 (1977).
\bibitem{weinberg} S.~Weinberg, Phys.\ Rev.\ Lett.\ {\bf 40}, 223 (1978).
\bibitem{wilczek} F.~Wilczek, Phys.\ Rev.\ Lett.\ {\bf 40}, 279 (1978).
\bibitem{griest} G.~Jungman, M.~Kamionkowski, K.~Griest, Phys.\ Rept.\ {\bf 267}, 195 (1996); G.~Bertone, D.~Hooper, J.~Silk  Phys.\ Rept.\ {\bf 405}, 279 (2005); L.~Bergstrom, New J.\ Phys.\ {\bf 11}, 105006 (2009).
\bibitem{barrow} J.~D.~Barrow, Nucl.\ Phys.\  B {\bf 208}, 501 (1982).
\bibitem{preskill} J.~Preskill, M.~Wise, F. Wilczek, Phys.\ Lett.\ B {\bf 120}, 127 (1983); L.~Abbott, P.~Sikivie, {\it ibid.}, 133; F.~W.~Stecker, Q.~Shafi, Phys.\ Rev.\ Lett. {\bf 50}, 928 (1983).
\bibitem{komatsu} E.~Komatsu {\it et al.} [WMAP Collaboration], Astrophys.\ J.\ Suppl.\ {\bf 180}, 330 (2009).
\bibitem{lazarides} G.~Lazarides, C.~Panagiotakopoulos, Q.~Shafi, Phys.\ Lett.\ B {\bf 192}, 323 (1987).
\bibitem{yamamoto} K.~Yamamoto, Phys.\ Lett.\ B {\bf 161}, 289 (1985).
\bibitem{donald} J.~McDonald, Phys.\ Rev.\ D {\bf 43}, 1063 (1991).
\bibitem{moroi} T.~Moroi, M.~Yamaguchi, T.~Yanagida, Phys.\ Lett.\ B {\bf 342}, 105 (1995); M.~Kawasaki, T.~Moroi, T.~Yanagida, Phys.\ Lett.\ B {\bf
    370}, 52 (1996).
\bibitem{kawasaki} M.~Kawasaki, T.~Moroi, T.~Yanagida, Phys.\ Lett.\ B {\bf 383}, 313 (1996).
\bibitem{chung} D.~J.~Chung, E.~W.~Kolb, A.~Riotto, Phys.\ Rev.\ D {\bf  60}, 063504 (1999).
\bibitem{riotto} G.~F.~Giudice, E.~W.~Kolb, A.~Riotto, Phys.\ Rev.\ {\bf D} 64, 023508 (2001).
\bibitem{giudice} G.~F.~Giudice, E.~W.~Kolb, A.~Riotto, D.~V.~Semikoz, I.~I.~Tkachev, Phys.\ Rev.\ D {\bf 64}, 043512 (2001).
\bibitem{gelmini} G.~Gelmini, S.~Palomares-Ruiz, S.~Pascoli, Phys.\ Rev.\ Lett.\ {\bf 93}, 081302 (2004).
\bibitem{gelmini1} G.~Gelmini, P.~Gondolo, Phys.\ Rev.\ D {\bf 74}, 023510 (2006).
\bibitem{drees} M.~Drees, H.~Iminniyaz, M.~Kakizaki, Phys.\ Rev.\ D {\bf 73}, 123502 (2006).
\bibitem{gelmini2} G.~Gelmini, P.~Gondolo, A.~Soldatenko, C.~E.~Yaguna, Phys.\ Rev.~ D {\bf 74}, 083514 (2006).
\bibitem{ford} L.~H.~Ford, Phys.\ Rev.\ D {\bf 35}, 2955 (1987).
\bibitem{kamionkowski_turner} M.~Kamionkowski, M.~S.~Turner, Phys.\ Rev.\ D {\bf 42}, 3310 (1990).
\bibitem{spokoiny} B.~Spokoiny, Phys.\ Lett.\ B {\bf 315}, 40 (1993).
\bibitem{joyce} M.~Joyce, Phys.\ Rev.\ D {\bf 55}, 1875 (1997); M.~Joyce, T.~Prokopec, Phys.\ Rev.\ D {\bf 57}, 6022 (1998).
\bibitem{tashiro} H.~Tashiro, T.~Chiba, M.~Sasaki, Class.\ Quant.\ Grav.\ {\bf 21}, 1761 (2004).
\bibitem{salati} P.~Salati, Phys.\ Lett.\ B {\bf 571}, 121 (2003).
\bibitem{profumo} S.~Profumo, P.~Ullio, JCAP {\bf 0311}, 006 (2003).
\bibitem{rosati} F.~Rosati, Phys.\ Lett.\ B {\bf 570}, 5 (2003).
\bibitem{pallis} C.~Pallis, JCAP {\bf 0510}, 015 (2005); Nucl.\ Phys.\ B {\bf 751}, 129 (2006).
\bibitem{chun} E.~J.~Chun, S.~Scopel, JCAP {\bf 0710}, 011 (2007).
\bibitem{chung1} D.~J.~Chung, L.~L.~Everett, K.~Kong, K.~T.~Matchev, JHEP {\bf 0710}, 016 (2007); Phys.\ Rev.\ D {\bf 76}, 103530 (2007).
\bibitem{gomez}  M.~E.~Gomez, S.~Lola, C.~Pallis, J.~Rodriguez-Quintero, JCAP {\bf 0901}, 027 (2009).
\bibitem{kamionkowski} D.~Grin, T.~L.~Smith, M.~Kamionkowski, Phys.\ Rev.\ {\bf D} 77, 085020 (2008).
\bibitem{kolb} E.~W.~Kolb, M.~S.~Turner, {\it The Early Universe}, Addison-Wesley (1990).
\bibitem{raffelt} G.~G.~Raffelt, J.\ Phys.\ A {\bf 40}, 6607 (2007).
\bibitem{kim_review} J.~E.~Kim, G.~Carosi, arXiv:0807.3125v2.
\bibitem{KSVZ} J.~E.~Kim, Phys.\ Rev.\ Lett.\ {\bf 43}, 103 (1979); M.~A.~Shifman, A.~I.~Vainshtein, V.~I.~Zakharov, Nucl.\ Phys.\ B {\bf 166}, 493 (1980).
\bibitem{DFSZ}  A.~P.~Zhitnitskii, Sov.\ J.\ Nucl.\ Phys.\ {\bf 31}, 260 (1980); M.~Dine, W.~Fischler, M.~Srednicki, Phys.\ Lett.\ B {\bf 104}, 199
    (1981).
\bibitem{turner_theta} M.~S.~Turner, Phys.\ Rev.\ D {\bf 33}, 889 (1986).
\bibitem{davis} R.~Davis, Phys.\ Rev.\ D {\bf 32}, 3172 (1985); Phys. Lett. B {\bf 180}, 225 (1986).
\bibitem{harari} D.~Harari, P.~Sikivie, Phys.\ Lett.\ B {\bf 195}, 361 (1987).
\bibitem{turner} M.~S.~Turner, Phys.\ Rev.\ Lett.\ {\bf 59}, 2489 (1987) [Erratum {\it ibid.} {\bf 60}, 1101 (1988)].
\bibitem{battye} R.~A.~Battye, E.~P.~S.~Shellard, Nucl.\ Phys.\ B {\bf 423}, 260 (1994); Phys.\ Rev.\ Lett.\ {\bf 76}, 2203 (1996).
\bibitem{khlopov} A.~S.~Sakharov, M.~Y.~Khlopov, Phys.\ Atom.\ Nucl.\ {\bf 57}, 485 (1994) [Yad.\ Fiz.\ {\bf 57}, 514 (1994)]; A.~S.~Sakharov, D.~D.~Sokoloff, M.~Y.~Khlopov, Phys.\ Atom.\ Nucl.\ {\bf 59}, 1005 (1996) [Yad.\ Fiz.\ {\bf 59N6}, 1050 (1996)]; M~Y.~Khlopov, A.~S.~Sakharov, D.~D.~Sokoloff, Nucl.\ Phys.\ Proc.\ Suppl.\ {\bf 72}, 105 (1999).
\bibitem{hagmann} C.~Hagmann, S.~Chang, P.Sikivie, Nucl.\ Phys.\ B Proc.\ Suppl.\ {\bf 72}, 81 (1999).
\bibitem{shellard} E.~P.~S.~Shellard, R.~A.~Battye, Nucl.\ Phys.\ Proc.\ Suppl.\ {\bf 72}, 88 (1999).
\bibitem{wall_solution} S.~Chang, C.~Hagmann, P.~Sikivie, Phys.\ Rev.\ D {\bf 59}, 023505 (1999).
\bibitem{string_simulation} M.~Yamaguchi, M.~Kawasaki, J.~Yokoyama, Phys.\ Rev.\ Lett.\ {\bf 82}, 4578 (1999).
\bibitem{fox} P.~Fox, A.~Pierce, S.~Thomas, arXiv:hep-th/0409059v1
\bibitem{beltran} M.~Beltran, J.~Garcia-Bellido, J.~Lesgourgues, Phys.\ Rev.\ D {\bf 75}, 103507 (2007).
\bibitem{sikivie} P.~Sikivie, Lect.\ Notes Phys.\ {\bf 741}, 19-50 (2008).
\bibitem{hertzberg}  M.~P.~Hertzberg, M.~Tegmark, F.~Wilczek, Phys.\ Rev.\ D {\bf 78}, 083507 (2008).
\bibitem{bae} K.~J.~Bae, J.~Huh, J.~E.~Kim, JCAP {\bf 0809}, 005 (2008).
\bibitem{yang} P.~Sikivie, Q.~Yang, Phys.\ Rev.\ Lett.\ {\bf 103}, 111301 (2009).
\bibitem{hwang} J.~Hwang, H.~Noh,  Phys.\ Lett.\ B {\bf 680}, 1 (2009).
\bibitem{baer} H.~Baer, A.~D.~Box, H.~Summy, JHEP {\bf 0908}, 080 (2009).
\bibitem{visinelli} L.~Visinelli, P.~Gondolo, Phys.\ Rev.\ D {\bf 80}, 035024 (2009).
\bibitem{hamann} J.~Hamann, S.~Hannestad, G.~G.~Raffelt, Y.~Y.~Y.~Wong, JCAP {\bf 0906}, 22 (2009).
\bibitem{wantz} O.~Wantz, E.~P.~S.~Shellard, arXiv:0910.1066v2
\bibitem{mack} K.~J.~Mack, arXiv:0911.0421v1
\bibitem{gross} D.~J.~Gross, R.~D.~Pisarski, L.~G.~Yaffe, Rev.\ Mod.\ Phys.\ {\bf 53}, 43 (1981).
\bibitem{lyth} D.~H.~Lyth, Phys.\ Rev.\ D {\bf 45}, 3394 (1992).
\bibitem{strobl} K.~Strobl, T.~J.~Weiler, Phys.\ Rev.\ D {\bf 50}, 7690 (1994).
\bibitem{linde} A.~D.~Linde, Phys.\ Lett.\ B {\bf 201}, 437 (1988).
\bibitem{birrell} N.~D.~Birrell, P.~C.~W.~Davies, {\it Quantum Field Theory in Curved Space-Time}, Cambridge University Press (1982).
\bibitem{raffelt1} G.~G.~Raffelt, Lect.\ Notes\ Phys.\ {\bf 741}, 51 (2008).
\bibitem{vilenkin} A.~Vilenkin, A.~E.~Everett, Phys.\ Rev.\ Lett.\ {\bf 48}, 1867 (1982).
\bibitem{hagmann1} C.~Hagmann, P.~Sikivie, Nucl.\ Phys.\ B {\bf 363}, 247 (1991).
\bibitem{sikivie_wall} P.~Sikivie, Phys.\ Rev.\ Lett.\ {\bf 48}, 1156 (1982).
\bibitem{davis2} R.~L.~Davis, E.~P.~S.~Shellard, Nucl.\ Phys.\ B {\bf 324}, 167 (1989); A.~Dabholkar, J.~M.~Quashnock, Nucl.\ Phys.\ B {\bf 333}, 815 (1990).
\bibitem{local_strings_results} D.~P.~Bennett, F.~R.~Bouchet, Phys.\ Rev.\ D {\bf 41}, 2408 (1990); B.~Allen, E.~P.~S.~Shellard, Phys.\ Rev.\ Lett.\ {\bf 64}, 119 (1990).
\bibitem{string_matter} M.~Yamaguchi, J.~Yokoyama, M.~Kawasaki, Phys.\ Rev.\ D {\bf 61}, 061301 (2000).
\bibitem{asztalos} S.~Asztalos {\it et al.}, Phys.\ Rev.\ D {\bf 64} 092003 (2001); {\it ibid.} {\bf 69}, 011101(R) (2004).
\bibitem{duffy} L.~D.~Duffy {\it et al.}, Phys.\ Rev.\ D {\bf 74}, 012006 (2006).

\end{thebibliography}
\end{document}